\documentclass[journal]{IEEEtran}
\usepackage{cite}
\usepackage{color}

\ifCLASSINFOpdf
   \usepackage[pdftex]{graphicx}
   \DeclareGraphicsExtensions{.pdf,.jpeg,.png}
\else
  \usepackage[dvips]{graphicx}
  \DeclareGraphicsExtensions{.eps}
\fi
\usepackage[cmex10]{amsmath}
\usepackage{array}
\usepackage{amsmath}
\usepackage{url}
\usepackage{bm}
\begin{document}
\title{A resistance bridge based on a cryogenic current comparator achieving sub-$10^{-9}$ measurement uncertainties}


\author{Wilfrid Poirier, Dominique Leprat and F\'elicien Schopfer% <-this % stops a space
\thanks{Authors are with the Department of Fundamental Electrical Metrology,
Laboratoire national de m\'etrologie et d'essais, 78197 Trappes, France; e-mail: wilfrid.poirier@lne.fr.}}%
\maketitle

\begin{abstract}
A new resistance bridge has been built at the Laboratoire national de m\'etrologie et d'essais (LNE) to improve the ohm realization in the \emph{Syst\`eme International} (SI) of units from the quantum Hall effect. We describe the instrument, the performance of which relies on two synchronized and noise-filtered current sources, an accurate and stable current divider and a cryogenic current comparator (CCC) having a low noise of $\mathrm{80~pA.t/Hz^{1/2}}$. The uncertainty budget for the measurement of the 100 $\Omega/(R_\mathrm{K}/2)$ ratio, where $R_\mathrm{K}$ is the von Klitzing constant, amounts to a few parts in $10^{10}$ only.
\end{abstract}


% \fi
\IEEEpeerreviewmaketitle
\section{Introduction}
\IEEEPARstart{I}{n} the SI\cite{BIPM}, the ohm can be realized from $R_\mathrm{K}=h/e^2$\cite{AmpereBIPM2019}, where $h$ is the Planck constant, $e$ is the elementary charge, using the quantum Hall effect. In national metrology institutes (NMIs), the quantized Hall resistance, $R_\mathrm{K}/i$, where $i$ is an integer, is used as an universal primary reference to disseminate the ohm by means of resistance comparisons\cite{Poirier2019}. Performing these resistance comparisons is challenging since the measurement current of the quantum Hall resistance (QHR) devices must remain small, i.e. a few tens of $\mu$A if based on GaAs/AlGaAs heterostructure\cite{Jeckelmann2001,Poirier2009} and a few hundreds of $\mu$A if based on graphene\cite{Ribeiro2015}. The most accurate and sensitive resistance bridge (RB), based on the performance of a cryogenic current comparator (CCC), is able to achieve relative measurements uncertainties of a few $10^{-9}$. The CCC\cite{Harvey1972} is basically a perfect transformer operating in direct current regime (dc) able to measure the ratio of the currents circulating through the two resistances to compare with a relative uncertainty below $10^{-10}$. Made of superconducting windings embedded in a superconducting shielding, its accuracy relies on the Meissner effect. Its high current sensitivity comes from the flux detector equipping it, which is based on a Superconducting QUantum Interference Device (SQUID)\cite{Gallop2006}.

The development of resistance bridge equipped with a CCC started in several national metrology institutes (NMI), including the French institute, right after the discovery of the QHE. First ones were equipped with radio-frequency (rf) SQUID and used with dc\cite{Delayahe1985,Williams1991,Hartland92,Dziuba1993}. The LNE has been using such a dc bridge\cite{Delayahe1985,Piquemal1995} for more than thirty years to perform calibrations of wire resistors with a relative measurement uncertainty of a few $10^{-9}$, as more than twenty NMIs do at the present time. Other bridges adapted to measurements in the low-frequency (below 1 Hz) alternating current regime (ac) were then proposed\cite{Delahaye1991,Seppa1997}. Accurate operation at higher frequencies was achieved by replacing the CCC with a room-temperature current comparator using high-permeability magnetic cores\cite{DelahayeAC1993,Satrapinski2017} but at the expense of larger measurement uncertainties. The improvement of digital and analog electronic components and the availability of dc SQUIDs then allowed several NMIs to develop resistance bridges with better performance in terms of sensitivity, accuracy and automation\cite{Sanchez2009,Drung2009,Gotz2009,Williams2010}. Despite this, achieving a combined measurement uncertainty below $10^{-9}$ remains very challenging. Among the NMIs having participated to the International bilateral comparisons\cite{BIPMEMK12} with the \emph{Bureau International des Poids et Mesures} (BIPM), which are considered as giving one of the best validation of measurement uncertainty budgets, only three have reported on such low measurement uncertainties.

Here, we report on a new RB based on dc SQUID designed to perform comparisons of resistances over a wider range of values (from $1~\Omega$ to $\mathrm{1.29~M\Omega}$) and with lower measurement uncertainties than with the older bridge. Our estimated combined standard uncertainty for the measurement of the 100 $\Omega/(R_\mathrm{K}/2)$ ratio amounts to $0.6\times10^{-9}$ ($1\sigma$) only, in relative value. The type A relative uncertainty achieved for one hour measurement can be as low as $0.2\times10^{-9}$. This compares with the five times larger uncertainty obtained with the older bridge due to the outdated performance of its rf SQUID.
\section{Principle of the RB}
\begin{figure}[h]
\centering
\includegraphics[width=3.5in]{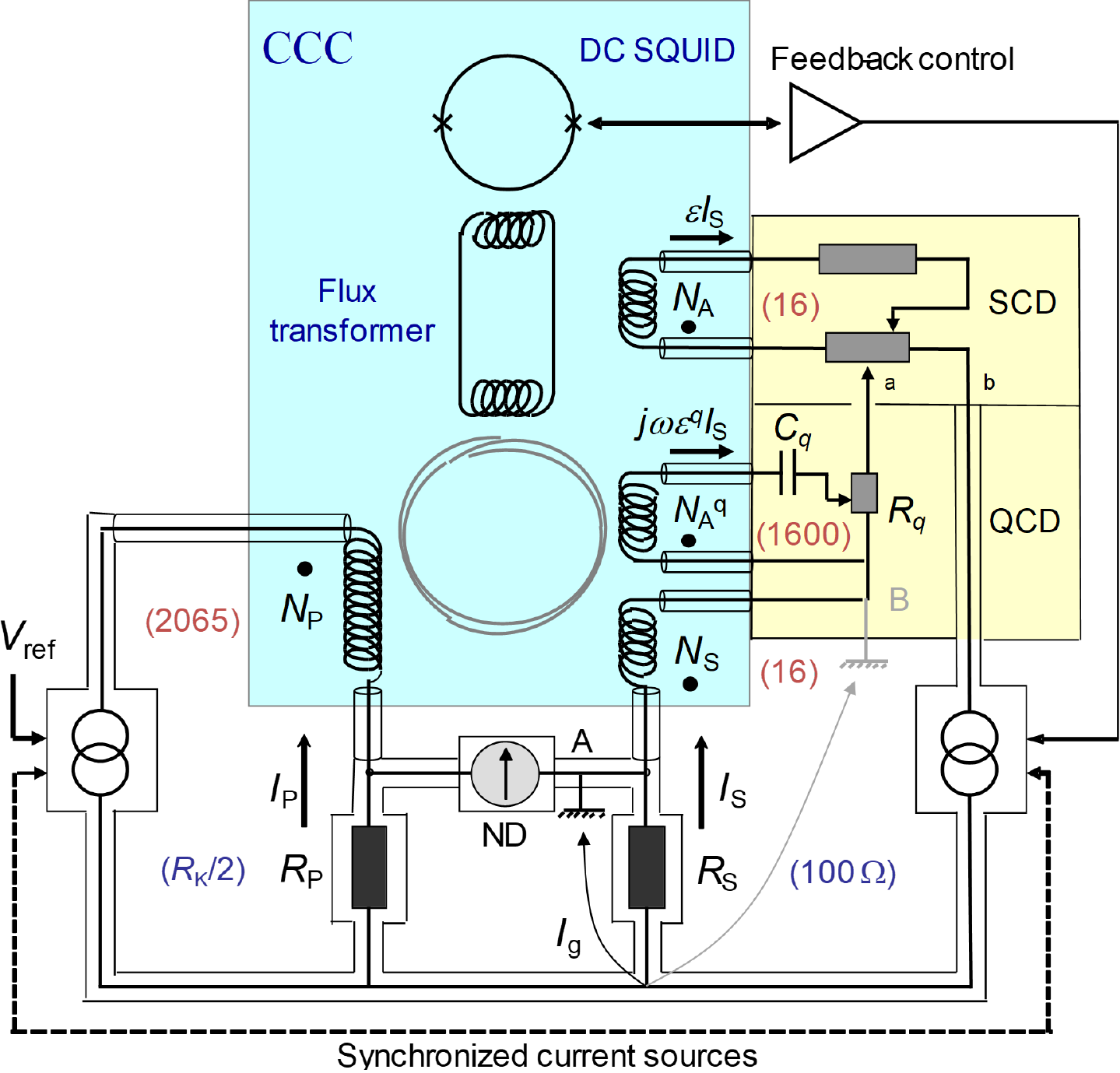}
\caption{Principle of the new LNE resistance bridge based on a CCC. The figure shows the two synchronized current sources, the CCC equipped with a dc SQUID and the feedback control on the secondary current source, the standard (SCD) and the quadrature (QCD) current dividers, the null detector (ND) and the two resistances $R_\mathrm{S}$ and $R_\mathrm{P}$ to compare. The ground can be connected in position A (low potential of the secondary resistor) or B (low potential of the secondary winding).}\label{fig1}
\end{figure}
The principle of the new RB, described in fig.\ref{fig1}, is close to that of the older one. It is based on two synchronized sources, primary (P) and secondary (S) sources, that deliver currents $I_\mathrm{P}$ and $I_\mathrm{S}$ respectively. The primary (secondary) source supplies the resistance $R_\mathrm{P}$ ($R_\mathrm{S}$) in series with a superconducting winding of a CCC of number of turns $N_\mathrm{P}$ ($N_\mathrm{S}$). The number of turns $N_\mathrm{S}$ and $N_\mathrm{P}$ are chosen so that the ratio $N_\mathrm{S}/N_\mathrm{P}$ is close to the resistance ratio $R_\mathrm{S}/R_\mathrm{P}$. A standard current divider (SCD) is used to deviate an in-phase calibrated fraction $\epsilon$ of the current $I_\mathrm{S}$ into an auxiliary winding of number of turns $N_\mathrm{A}$. The windings of the CCC are wound according to a toroidal geometry and embedded in a superconducting shielding. Application of the Ampere's theorem to a circulation along a cross-section of the shielding, where the magnetic flux density is zero, leads to the relationship $N_\mathrm{P}I_\mathrm{P}-(N_\mathrm{S}+\epsilon N_\mathrm{A})I_\mathrm{S}=I_\mathrm{CCC}$, where $I_\mathrm{CCC}$ is a screening current. Because the CCC shielding overlaps itself without electrical contact, this superconducting current circulates from the inner to the outer side of the shielding. It is detected by a pick-up coil coupled to the outer side and connected to the entry inductance of a dc SQUID. The secondary current source is servo-controlled by the output of the CCC SQUID electronics so that the screening current $I_\mathrm{CCC}$ (i.e. the total ampere.turn) is nulled. It results that:
\begin{equation}
N_\mathrm{P}I_\mathrm{P}-(N_\mathrm{S}+\epsilon N_\mathrm{A})I_\mathrm{S}=0.
\label{Equation:AmpereTurns}
\end{equation}
From the fraction $\epsilon_\mathrm{eq}$ setting the voltage balance (equilibrium), $R_\mathrm{S}I_\mathrm{S}=R_\mathrm{P}I_\mathrm{P}$, one obtains:
\begin{equation}
R_\mathrm{S}/R_\mathrm{P}=(N_\mathrm{S}+\epsilon_\mathrm{eq} N_\mathrm{A})/N_\mathrm{P}.
\label{Equation:resistance}
\end{equation}
The SCD can also be inserted in the primary circuit to deviate a fraction of the current $I_\mathrm{P}$. In this case, the previous equations remain valid by simply exchanging S and P index. This is the operating mode planned for measurements involving a low resistance $R_\mathrm{S}$ (for example 1 $\Omega$) supplied by a large current $I_\mathrm{S}$ (for example 50 mA) which would lead to a too strong dissipation in the SCD if placed in the secondary circuit. Instead, the SCD inserted in the primary circuit is biased by the lower current $I_\mathrm{P}$ which is usually below 10 mA.

Compared to the older RB, main improvements implemented in the new bridge concern i) the two current sources which are carefully noise-filtered and finely synchronized by a single external voltage source so that they are able to operate both in dc (square current signal with periodic current reversal) and in ac at very low frequencies, ii) the new dc SQUID-based CCC which is equipped with more windings and is characterized by a lower noise level, iii) the standard current divider which is more accurate and stable, iv) the shielding which has a better continuity and contributes to lower noise and better accuracy, and v) the addition of a quadrature current divider (QCD) used to cancel the voltage overshoots (or in-quadrature signal) caused by stray capacitances which occur at the entry of the null detector (ND) during current reversals (or ac current variation).

Voltage overshoots, which become large for short current rise time, can lead to saturation of the ND during current transitions. To avoid this, one can adopt long enough current rise time and large filter time constant for the ND. But, this limits the current reversal frequency which is not favorable to offset subtraction and rejection of the 1/$f$ SQUID noise. The QCD offers an alternative solution applicable both in dc regime and in low-frequency ac regime. This new device can deviate an in-quadrature current fraction $j\omega\epsilon^{q}I_\mathrm{S}$ in a fourth winding of number of turns $N_\mathrm{A}^q$, where $\omega$ is the angular frequency, $\epsilon^{q}=\alpha R_qC_q$, and $\alpha\in[0:1]$ is a fraction adjusted by a potentiometer (fig.\ref{fig1}). The equation \ref{Equation:AmpereTurns} for ampere.turns becomes:
\begin{equation}
N_\mathrm{P}I_\mathrm{P}-(N_\mathrm{S}+\epsilon N_\mathrm{A}+j\omega\epsilon^{q}N_\mathrm{A}^q)I_\mathrm{S}=0.
\label{Equation:AmpereTurnsQCD}
\end{equation}
Considering stray capacitances $C_\mathrm{P}$ and $C_\mathrm{S}$ in parallel to the resistors $R_\mathrm{P}$ and $R_\mathrm{S}$ respectively and assuming first-order approximation in $\omega$, the voltage balance is achieved provided that the fraction $\epsilon_\mathrm{eq}$ obeys the equation \ref{Equation:resistance} and that the fraction $\epsilon^q_\mathrm{eq}$ fulfills the equation (see Appendix \ref{appendix:QCD}):
\begin{equation}
(R_\mathrm{P}C_\mathrm{P}-R_\mathrm{S}C_\mathrm{S})\omega=\epsilon^{q}_\mathrm{eq}\omega \frac{N_\mathrm{A}^q}{(N_\mathrm{S}+\epsilon_\mathrm{eq}N_\mathrm{A})}\simeq \epsilon^{q}_\mathrm{eq}\omega\frac{N_\mathrm{A}^q}{N_\mathrm{S}}.
\label{Equation:Phase}
\end{equation}
This equation emphasizes that the in-quadrature voltage signal due to stray capacitances can be compensated by the injection of the in-quadrature current, for any angular frequency $\omega$, by setting the QCD fraction to the value $\epsilon^{q}_\mathrm{eq}=\alpha_\mathrm{eq}R_qC_q$. In dc operation of the bridge, the QCD allows the cancellation of the in-quadrature signals at the harmonic frequencies of the current reversal frequency, which form the parasitic voltage overshoots. One can therefore decrease the ND filter time constant and the current rise time and also increase the current reversal frequency without any risk of saturation of the ND. This allows reducing the impact of the voltage offset drift and of the $1/f$ SQUID noise on measurements and increasing the ratio between the acquisition time and the total experience time. Let us remark that the calibration of the QCD fraction is not required.
\section{the CCC}
\subsection{Design and fabrication}
\begin{figure}[h]
\centering
\includegraphics[width=3.5in]{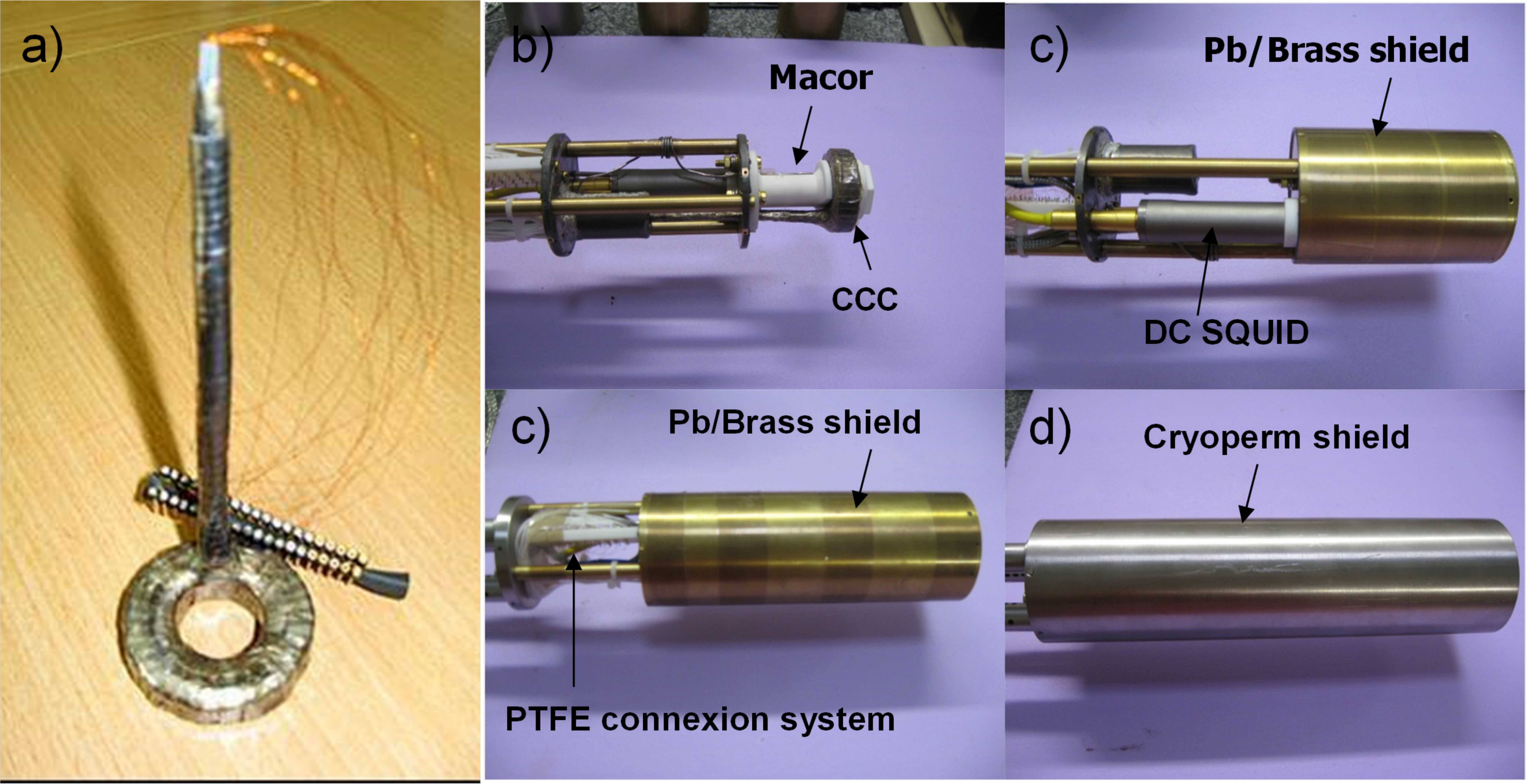}
\caption{Pictures of the CCC at different stages of shielding assembly. a) The CCC alone: the outer diameter, the inner diameter and the height of the CCC are 42 mm, 19 mm and 12 mm respectively. The length of the chimney is of 110 mm. b) The CCC on the probe with all shields removed. An additional superconducting cylindrical piece (length of 40 mm) extends the shielding of the chimney exit wires. c) First Pb/Brass shield (length/diameter: 88 mm/60 mm) of the CCC in place. d) Second Pb/Brass shield (length/diameter: 180 mm/64 mm) of the SQUID in place. e) External cryoperm shield (length/diameter: 256 mm/68 mm) in place.}\label{Fig-CCC}
\end{figure}
The cryogenic current comparator, shown in fig.\ref{Fig-CCC}a), is made of 15 windings of  1, 1, 2, 4, 16, 16, 32, 64, 128, 160, 160, 1600, 1600, 2065 and 2065 turns which are held together with epoxy glue\cite{Soukiassian2010}. Each winding is made of superconducing and insulated $\mathrm{60~\mu m}$ diameter NbTi/Cu wire. We used optically checked 0.1 mm thick Pb sheets and Pb/Sn/Cd superconducting solder at a temperature lower than the Pb melting temperature ($\sim150^{~\circ}$C) to realize the toroidal shielding around the windings. Our shield overlaps twice (3 layers) to prevent from flux leakage. Each layer is covered with PTFE (poly-tetra-fluoro-ethylene) tape for electrical insulation. The CCC is fixed to the cryogenic probe with a piece in MACOR$^{\scriptsize\textregistered}$ material (fig.\ref{Fig-CCC}b)). The dc SQUID (Quantum Design, Inc) has an input inductance of $L_i=1.8~\mu H$ and a nominal flux noise in flux-lock feedback mode of $\mathrm{3~\mu\phi_0/Hz^{1/2}}$ above 0.3 Hz. It is coupled to the CCC via a superconducting flux transformer made of a NbTi wire inserted in a lead tube. Due to geometrical constraints, the number of turns of the pick-up coil was reduced to $N_\mathrm{PC}=6$. The magnetic screen is made of 5 concentric cylinders: two Pb ones, each one embedded in a brass one (fig.\ref{Fig-CCC}c), d)), and an outer Cryoperm® cylinder (fig.\ref{Fig-CCC}e)). Each cylinder is closed at the top with the same material. The cryogenic probe body is made of three rods used to stabilize it mechanically and reduce the CCC vibrations that cause electrical noise in measurements.

From the SQUID sensitivity, the $N_\mathrm{PC}$ number of turns, the SQUID input inductance $L_i$ and the effective CCC self-inductance, $L^\mathrm{eff}_\mathrm{CCC}\sim 14$ nH, determined taking into account the proximity of the superconducting screen\cite{Sese1999,Sese2003}, one can calculate (see Appendix \ref{annexe:CCC}) an expected CCC sensitivity, $S_\mathrm{CCC}$, of $\mathrm{6~\mu A.t/\phi_0}$. This value is close to that measured experimentally, $S_\mathrm{CCC}=\mathrm{8~\mu A.t/\phi_0}$.
\subsection{Noise performance}
\begin{figure}[h!]
\centering
\includegraphics[width=3.5in]{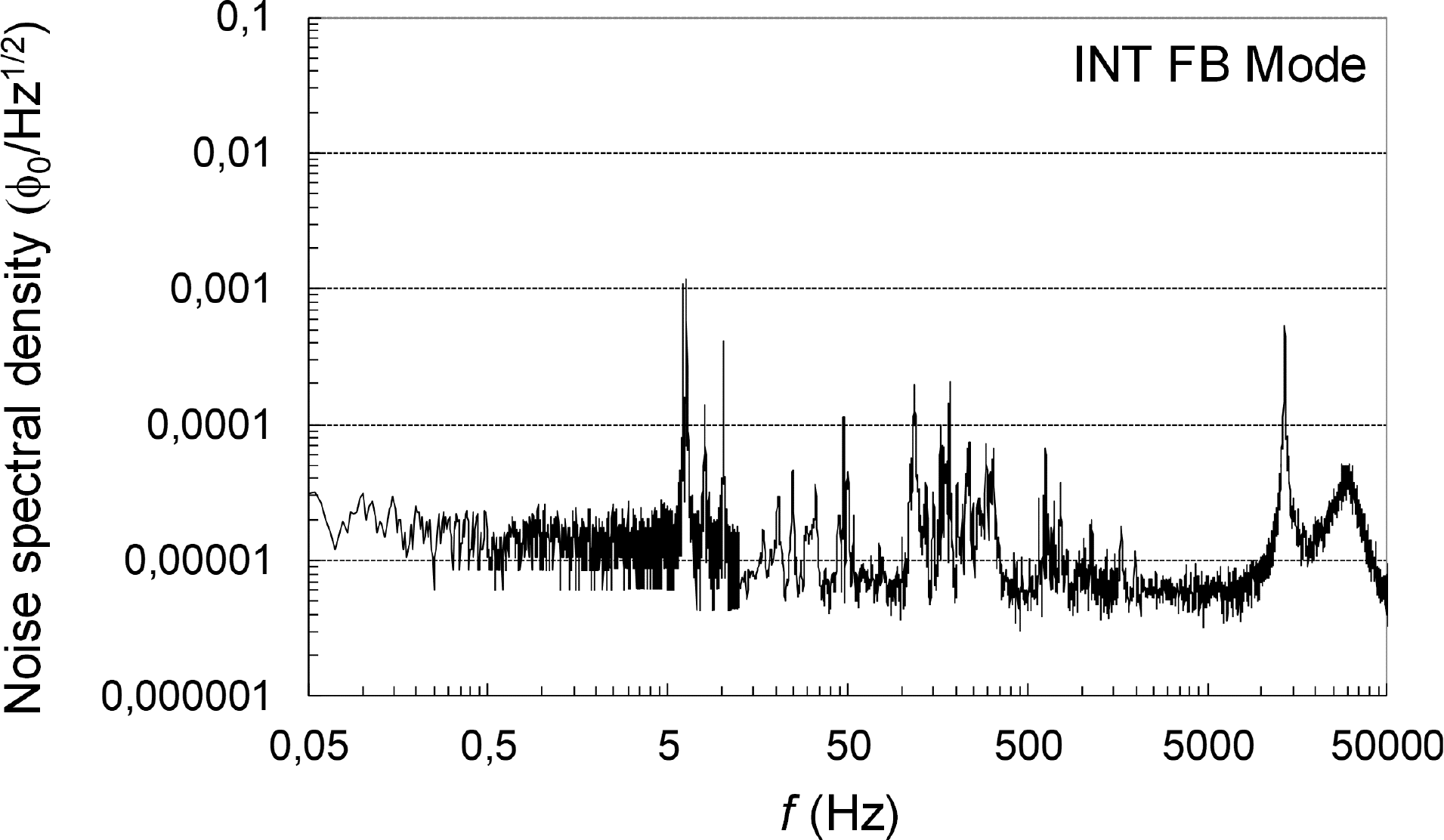}
\caption{Noise spectral density measured by the SQUID, operating in internal feedback mode (mode 5), \emph{versus} frequency \emph{f} for the CCC alone (no external cable connected to any winding)}.\label{Fig-NoiseCCCAlone}
\end{figure}
The CCC was firstly tested with all windings disconnected (no external cable connected at room temperature). Fig.\ref{Fig-NoiseCCCAlone} shows the noise spectral density in $\mathrm{\phi_0/Hz^{1/2}}$, measured by the SQUID operating in internal feedback mode (through the modulation coil of the SQUID) as a function of the frequency. The main frequency resonance due to the coupling of the large inductance of windings and the capacitance between wires is around 14 kHz. Between 6 Hz and about 2 kHz, the noise spectral density is dominated by sharp peaks with an amplitude lower than $1~\mathrm{m\phi_0/Hz^{1/2}}$ which are caused by mechanical resonances. At lower frequencies down to about 0.1 Hz, there exists a white noise regime with a constant noise spectral density of about $10~\mathrm{\mu \phi_0/Hz^{1/2}}$. Considering the CCC sensitivity of $\mathrm{8~\mu A.t/\phi_0}$, this leads to a current sensitivity of about $\mathrm{80~pA.t/Hz^{1/2}}$ which is fifteen times better than the $\mathrm{1300~pA.t/Hz^{1/2}}$ current sensitivity of the rf SQUID based CCC of the older bridge. At frequencies lower than 0.1 Hz, one can observe a noise increase which can be explained by the 1/\emph{f} noise of the Quantum Design dc SQUID. Operation of the resistance bridge with a current reversal frequency higher than 0.1 Hz should therefore lead to the lowest measurement noise.
\subsection{Accuracy}
Accuracy errors of the CCC are generally caused by a magnetic flux leakage which couples with the pick-up coil\cite{Jeckelmann2001,GallopHandbook2006}. They can be detected by supplying windings of same nominal number of turns $N$ connected in series-opposition with a large current $I$ ($=100$ mA, whatever $N$ value) and measuring the magnetic flux $\delta_{\phi_0}$ detected by the SQUID (operating in internal feedback mode). One then obtains the relative error $\Delta N/N= \delta_{\phi_0}S_\mathrm{CCC}/(NI)$, where $NI/S_\mathrm{CCC}$ should be the total flux generated by one winding. A low-noise current source is required for these measurements to reduce the SQUID errors caused by noise rectification. Best measurements were obtained using the current source of the RB (see subsection \ref{section:NoiseSpectrum}) and fig.\ref{Fig-NoiseCCCAutres}), which is characterized by a small frequency bandwidth of 1 kHz and is equipped with home-made common-mode torus reducing the circulation of noise from ground (see subsection \ref{Noise optimization and filtering}). For turn numbers \emph{N} equal to 16 and 2065 which are used in the calibration of a $100~\Omega$ resistor in terms of $R_\mathrm{K}/2$, $\Delta N/N$ is found equal to $(1.9 \pm 1.2)\times 10 ^{-11}$ and $(2.5 \pm 0.04)\times 10 ^{-11}$ respectively. For all other winding opposition, turn errors are found smaller than $6\times10^{-11}$, except for 1-1 and 2-2 combinations which seem let us conclude to significant errors of ~$\sim1\times10^{-9}$ and $\sim5\times10^{-10}$. One could explain this observation by a magnetic flux leakage from the wiring at the top of the chimney or a hole in the toroidal shield. However, the chimney is rather long and its end is well screened. Besides, a hole would also lead to errors for series-opposition of windings with a larger number of turns. Our interpretation is that these apparent larger errors are rather caused by spurious signals coming from residual noise rectification which manifest themselves all the more as the total ampere.turn number is small, i.e. for 1-1 and 2-2 winding opposition. Further reduction of the noise emitted by the current source in the 100 mA range (which is the most noisy range) is required to refine the determination of winding errors for small numbers of turns.
\begin{table}[h]
\newcolumntype{M}[1]{>{\centering\arraybackslash}m{#1}}
\begin{center}
\begin{tabular}{|c|M{5cm}|}
  \hline
  \textbf{Winding combination}& \textbf{$\Delta N/N$} \tabularnewline \hline
  1-1&$(1.29\pm 0.38)\times10^{-9}$\tabularnewline \hline
  2-2&$(4.64\pm 0.63)\times10^{-10}$ \tabularnewline \hline
  16-16&$(1.9\pm 1.2)\times10^{-11}$\tabularnewline \hline
  16+16-32&$(2.5\pm 1.1)\times10^{-11}$\tabularnewline \hline
  16+16+32-64&$(0.22\pm 0.76)\times10^{-11}$\tabularnewline \hline
  16+16+32+64-128&$(1.0\pm 1.1)\times10^{-11}$\tabularnewline \hline
  160-160&$(2.5\pm 0.14)\times10^{-11}$\tabularnewline \hline
  160-128-32&$(0.5\pm 0.4)\times10^{-11}$\tabularnewline \hline
  1600-1600&$(5.6\pm 0.032)\times10^{-11}$\tabularnewline \hline
  2065-2065&$(2.54\pm 0.035)\times10^{-11}$\\ \hline
\end{tabular}
\end{center}
\caption{Relative errors with standard uncertainties (k=1) of the number of turns, $\Delta N/N$, determined from winding opposition for different combinations of number of turns.}
\label{tableau:winding}
\end{table}

Our conclusion is that the CCC errors in the measurements of usual resistance ratios, which exploited mainly windings of number of turns 2065, 1600, 160, and 16, are of a few $10^{-11}$ only. These small residual errors are of similar magnitudes to those reported by other NMIs\cite{BIPMEMK12}. The CCC contribution to the type-B uncertainty, $u_\mathrm{B}^\mathrm{{CCC}}$, is therefore below $10^{-10}$, in relative value.
\section{The current source electronics}
\subsection{Design}
\begin{figure}[h!]
\centering
\includegraphics[width=3.5in]{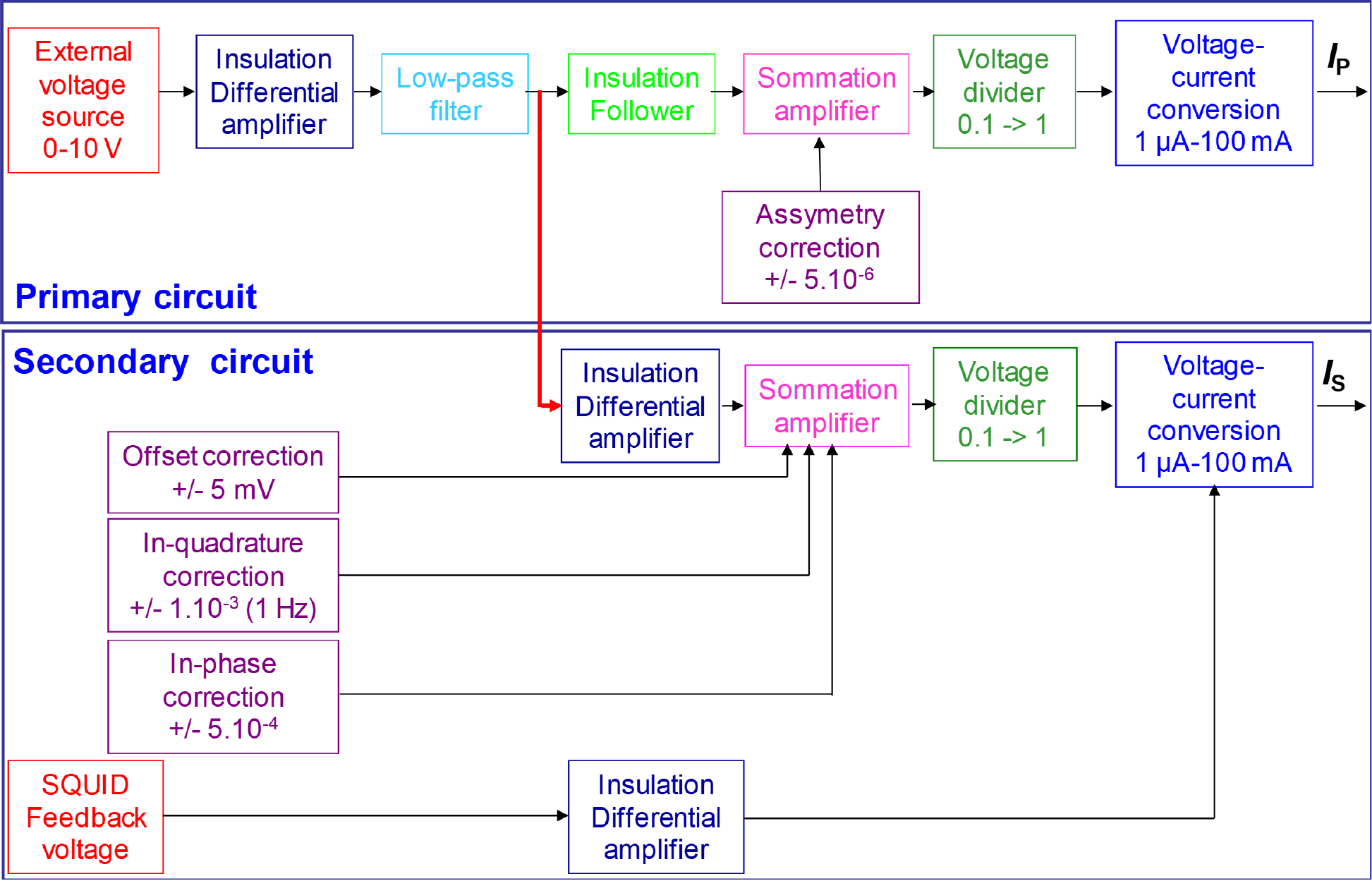}
\caption{Schemes of the electronic circuits of primary and secondary current sources. Both current sources are controlled by a single external voltage source.}\label{Fig-Circuits}
\end{figure}
The RB is based on two current sources generating the currents $I_\mathrm{P}$ and $I_\mathrm{S}$ which supply the resistors $R_\mathrm{P}$ and $R_\mathrm{S}$ respectively. In some recently developed resistance bridges, current sources are based on digital electronics connected by fiber optics to a internal micro-controller\cite{Drung2009,Williams2010} or an external PXI computer. This provides strong electrical insulation and easy automation but requires the implementation of efficient noise filtering techniques to protect the SQUID from the radio-frequency noise emitted by digital circuits.

On contrary, current sources of the new LNE RB remain based on linear analog circuits to avoid high-frequency noise\cite{Soukiassian2010}. Fig.\ref{Fig-Circuits} shows the schematic of the two electronic circuits, the primary and the secondary ones. They are controlled and synchronized by a single external voltage source allowing automation of measurements. The latter can be either a dc voltage generator, like a Yogogawa 7651 for usual dc measurements or the oscillator of a lock-in detector for low-frequency ac measurements. The external reference voltage supplies the primary circuit through a high-impedance differential amplifier. A low-pass filter is then used to limit the signal bandwidth with an adjustable cutoff frequency ranging from 1 mHz to 1 kHz. After a stage summing additional voltage corrections and a division stage allowing the setting of a decimal fraction, the voltage is converted into a current in ranges extending from 1 $\mu$A up to 100 mA. This conversion is done using an amplifier inverter circuit boosted by a buffer amplifier (BUF634T) and a dividing resistor $R_\mathrm{C}$ (see fig.\ref{Fig-Filtering}). The secondary current source, controlled by the output signal of the low-pass filter of the primary current source, is similar but the current range selected can also be multiplied by a factor 1.2906 or 1/1.2906 to adapt to the measurement of the specific resistance ratios involving the QHR connected either to the primary or to the secondary current sources. Besides, the design is such that there is no common mode voltage between the primary and secondary circuits (the 0 V reference of the two circuits, although electrically isolated, are at the same potential by design) which is beneficial to adjust the current ratio. Nevertheless, several additional circuits are required to finely adjust the current ratio $r_I=I_\mathrm{S}/I_\mathrm{P}$ to within a few parts in $10^{6}$ prior to the SQUID feedback operation. This is necessary to limit the ampere.turn unbalance in the CCC, not only to avoid unlocking of the SQUID feedback notably during current switching, but also to achieve the best accuracy in the $I_\mathrm{S}/I_\mathrm{P}$ ratio adjustment. Offset, in-phase and in-quadrature correction circuits are used to tune the secondary current. An asymmetry correction circuit, which corrects the main voltage by a small fraction of its absolute value, is also used in the primary circuit. It compensates, to some extent (to within a few $\times10^{-7}$), the asymmetry behaviour of the electronic circuits which manifests by a small change of the current ratio (typically $2\times10^{-6}$ in relative value) when reversing the current. Finally, the SQUID feedback voltage, after insulation using another high-impedance differential amplifier, is converted into a feedback current at the last stage of the secondary circuit so that the external feedback through the CCC winding has the same closed-loop gain as in internal feedback mode, i.e. $\sim0.75~V/\phi_0$.

Electronic circuits of current sources are made of precision operational amplifiers, high-stability and low temperature coefficient Vishay resistors. They are carefully shielded and electrically isolated from ground (see Appendix \ref{annexe:CurrentSourcesImplementation}).
\subsection{Test of the current ratio adjustability}
\label{subsection: Adjustability}
\begin{figure}[h!]
\centering
\includegraphics[width=3.5in]{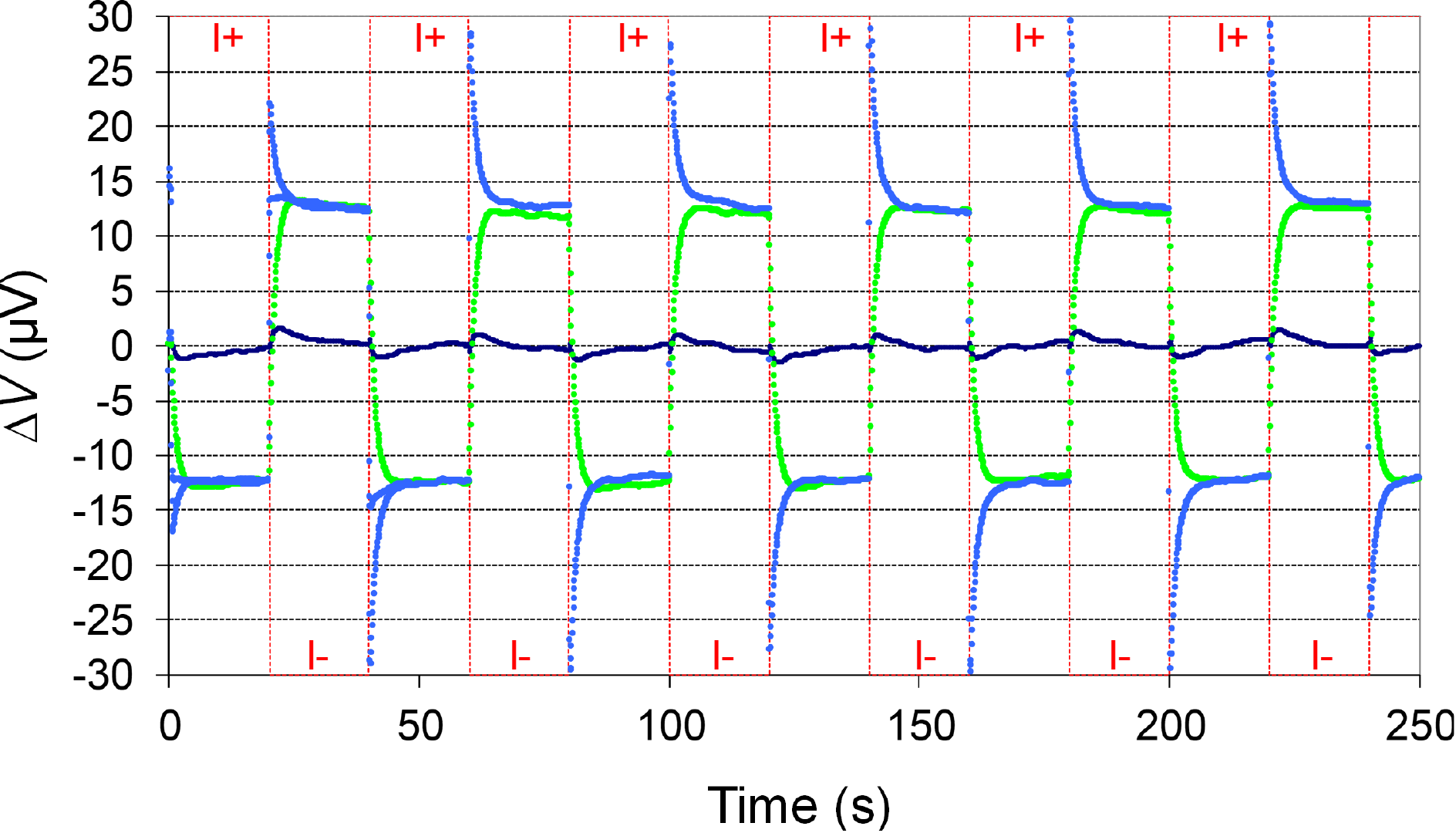}
\caption{Illustration of the adjustability of the current ratio $I_\mathrm{P}/I_\mathrm{S}$ using resistances $R_\mathrm{P}=\mathrm{10~k\Omega}$ and $R_\mathrm{S}=\mathrm{100~\Omega}$ (the SQUID feedback control is not operating). The balance voltage $\Delta V$, measured by the null detector (ND) is recorded as a function of time while the current is periodically reversed (red dashed line) for different settings of the corrections circuits : adjustment of the offset correction only (blue line), adjustment of the in-quadrature correction (green line), adjustment of in-phase, in-quadrature and asymmetry corrections (deep blue line).}\label{Fig-CurrentAdjustment}
\end{figure}
Fig.\ref{Fig-CurrentAdjustment} shows the experiment carried out to test the adjustability of the ratio of the two current sources which can be achieved prior to operating the SQUID feedback control\cite{Soukiassian2010}. Two resistors of resistance $\mathrm{10~k\Omega}$ and $\mathrm{100~\Omega}$ are fed by currents $I_\mathrm{P}$ and $I_\mathrm{S}$ respectively. The potential drop difference $\Delta V$ at the terminals of the two resistors is recorded by a null detector (nanovoltmeter EMN 11). For a nominal voltage reversing from 1 V to -1 V every 20 seconds, Fig.\ref{Fig-CurrentAdjustment} shows that it is possible to reduce the peak to peak $\Delta V$ amplitude to less than $\mathrm{2~\mu V}$ by adjusting the in-phase, the in-quadrature, the offset and the asymmetry corrections (the latter correction is used to symmetrize $\Delta V$ when reversing the current). Let us notably remark the effect of the in-quadrature correction in cancelling the voltage overshoot caused by the fast current reversal. This experiment proves that the current ratio can be adjusted to within $2\times10^{-6}$, in relative value, even during fast current reversal, which is an advantage both to avoid any SQUID unlocking and achieve best accuracy (see Appendix \ref{annexe:CLG}).
%\IEEEpubidadjcol
\subsection{Noise optimization and filtering}
\label{Noise optimization and filtering}
\begin{figure}[h!]
\centering
\includegraphics[width=3.5in]{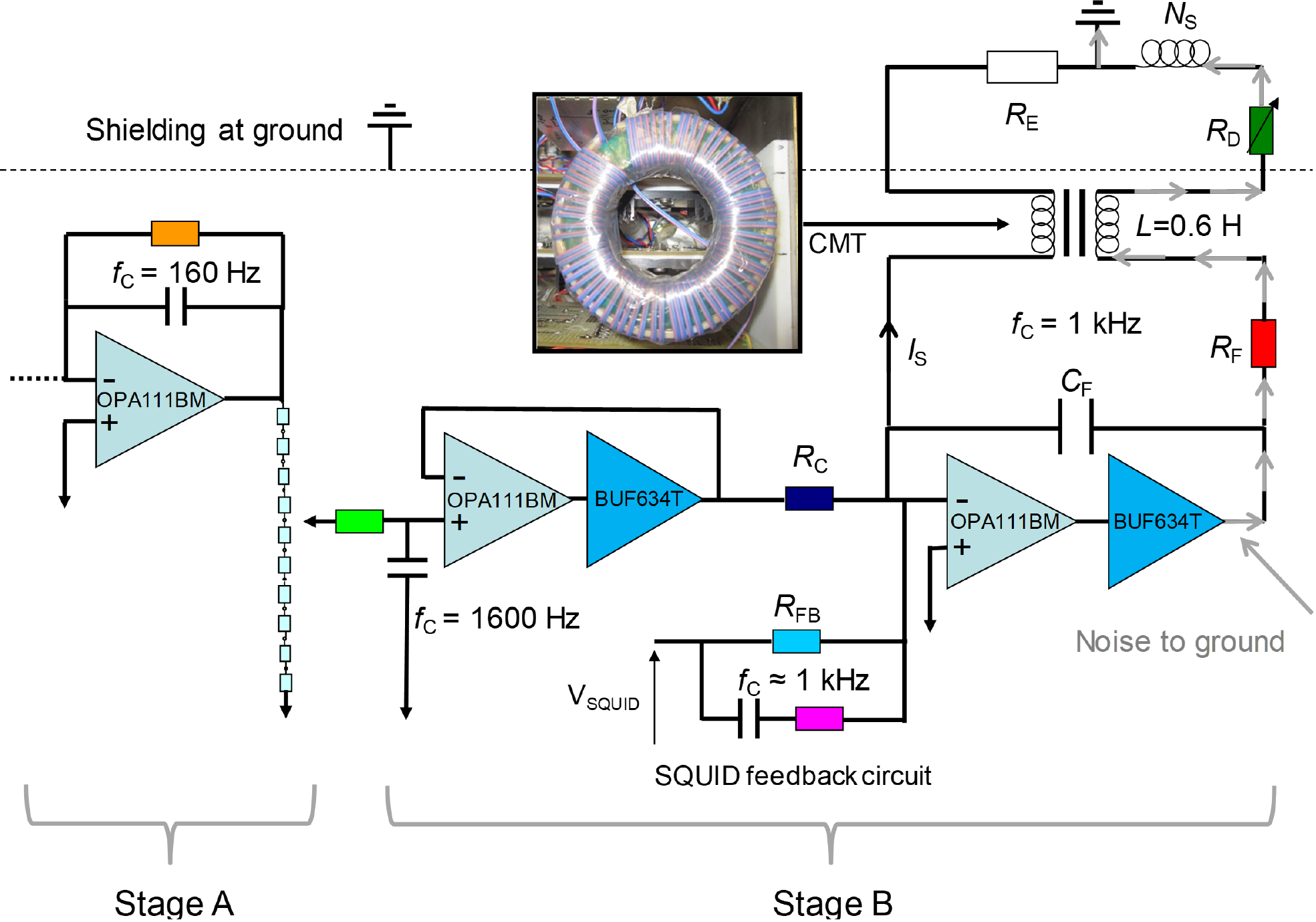}
\caption{Electronic scheme of the two last stages (A and B) of the secondary current source (the primary current source does not include the SQUID feedback circuit) describing the noise filtering techniques. Picture of the common mode torus (CMT) used to block the circulation of current noise towards ground (an example is shown with grey arrows). The dotted line represents the limit of the case (at ground) of the current source.}\label{Fig-Filtering}
\end{figure}
Noise filtering is crucial particularly from the resonance frequency of the CCC (14 kHz) up to the operating frequency of the modulation circuit (500 kHz) not only to ensure a good working of the SQUID but also to avoid noise rectification that would alter measurement accuracy. Fig.\ref{Fig-Filtering} shows the last stage of the electronic circuit of the secondary current source that supplies the resistance under measurement, $R_\mathrm{E}$, in series with a CCC winding. The primary current source is based on a similar stage but differs by the absence of the SQUID feedback electronic circuit. In practice, the frequency bandwidth of the primary and secondary current sources was reduced to 160 Hz using a simple low-pass filter at the stage A of the circuit. The filtered voltage is then converted into a current using the resistance $R_\mathrm{C}$, the value of which defines the current range. A second low-pass filter with a cutoff frequency of 1 kHz, defined by the capacitance $C_\mathrm{F}$, the resistances $R_\mathrm{F}$ and $R_\mathrm{E}$, is implemented at stage B to damp the CCC resonances. This filter limits the frequency bandwidth of the SQUID feedback circuit servo-controlling the secondary current to 1 kHz. Values of the passive components chosen for the different current ranges are summarized in Table.\ref{tableau:Range} of Appendix\ref{annexe:PassiveComponents}.

It is also essential to avoid the circulation of the current noise coming from the capacitive coupling of the electronics circuit with ground. To cancel this noise source which renders the SQUID inoperative, a home-made common mode torus (CMT) was introduced in the current circuit of each source (see fig.\ref{Fig-Filtering}). This CMT is made of a PTFE insulated wire pair wounded about 60 times around an APERAM Nano magnetic torus (magnetic permeability of about 80000 up to a 100 kHz frequency) with an anti-progression turn returning to the beginning of the winding\cite{Awan2011Book}. The differential inductance is around $\mathrm{3~\mu H}$ while the common mode inductance is around 0.6 H. The common mode impedance, which increases from about $200~\Omega$ at 50 Hz up to $\mathrm{150~k\Omega}$ at 1 MHz, drastically reduces the circulation of the common mode current noise. This protects the SQUID and makes it operating quite ideally whether it is a rf or a dc SQUID.
\subsection{The SQUID feedback circuit}
\label{Subsection:Feedback}
The Quantum Design SQUID can operate in four internal feedback operation modes, 5/5s, 50 and 500 of close-loop feedback gains $0.75~V/\phi_0$, $0.075~V/\phi_0$ and $0.0075~V/\phi_0$ respectively. These gains are defined by feedback resistors of resistance values, 500 k$\Omega$, 50 k$\Omega$ and 5 k$\Omega$ respectively, that divide the SQUID voltage detected after amplification by an integrator to generate the feedback current injected in the modulation coil. The SQUID bandwidth, defined by the characteristic frequency of the integrator amplifier (which defines the open-loop gain), is of 50 kHz except for the 5s mode for which it is reduced to 500 Hz. In external feedback mode, the feedback circuit is disconnected from the modulation coil and the voltage signal, $V_\mathrm{SQUID}$, detected at the output of the SQUID preamplifier, is sent to the secondary current source of the bridge after decoupling by a high-impedance differential amplifier. The SQUID feedback resistors therefore no more define the close-loop gains. On the other hand, another resistor $R_\mathrm{FB}$ biased by the $V_\mathrm{SQUID}$ voltage is used to define the feedback current supplying the secondary winding ($N_\mathrm{S}$). The resistor value, $R_\mathrm{FB}=\mathrm{1.5~M\Omega}$, is chosen so that the closed-loop feedback gain $G_\mathrm{CLG}=R_\mathrm{FB}\times S_\mathrm{CCC}/N_\mathrm{S}$ for $N_\mathrm{S}=16$ has the same value as in the most sensitive internal feedback mode of the SQUID (mode 5/5s), i.e. $0.75~V/\phi_0$. The stability of the closed-loop feedback operation also requires that the feedback circuit generates small signal dephasing. To partially compensate the dephasing caused by the 1 kHz low-pass filter implemented in the stage B of the electronic circuit and therefore optimize the SQUID operation, a circuit made of a small capacitance of 200 pF in series with a $\mathrm{20~k\Omega}$ resistor was connected in parallel to the $\mathrm{1.5~M\Omega}$ resistor.
\section{Current dividers}
\subsection{The standard (or in-phase) current divider}
The standard current divider (SCD) is used to balance the dc (or in-phase) voltages measured at the terminals of the two resistors by deviating a fraction of current towards the auxiliary winding. This is a key element, the accuracy of which directly impacts the uncertainty budget of the RB. An alternative technique to null the voltage measured by the detector consists in using a calibrated auxiliary current source servo-controlled by the null detector voltage output\cite{Williams1991,Hartland92}. On contrary, the SCD developed is a passive component which avoids the use of a second feedback electronics and aims for a good stability of all current fractions. This can be achieved with a design that limits the number of electrical commutations required to select the current fraction. The counterpart is that the calibration of the LNE SCD is not fully-automated contrary to that of binary compensation units\cite{Drung2013,Gotz2017}.
\begin{figure}[h]
\centering
\includegraphics[width=3.5in]{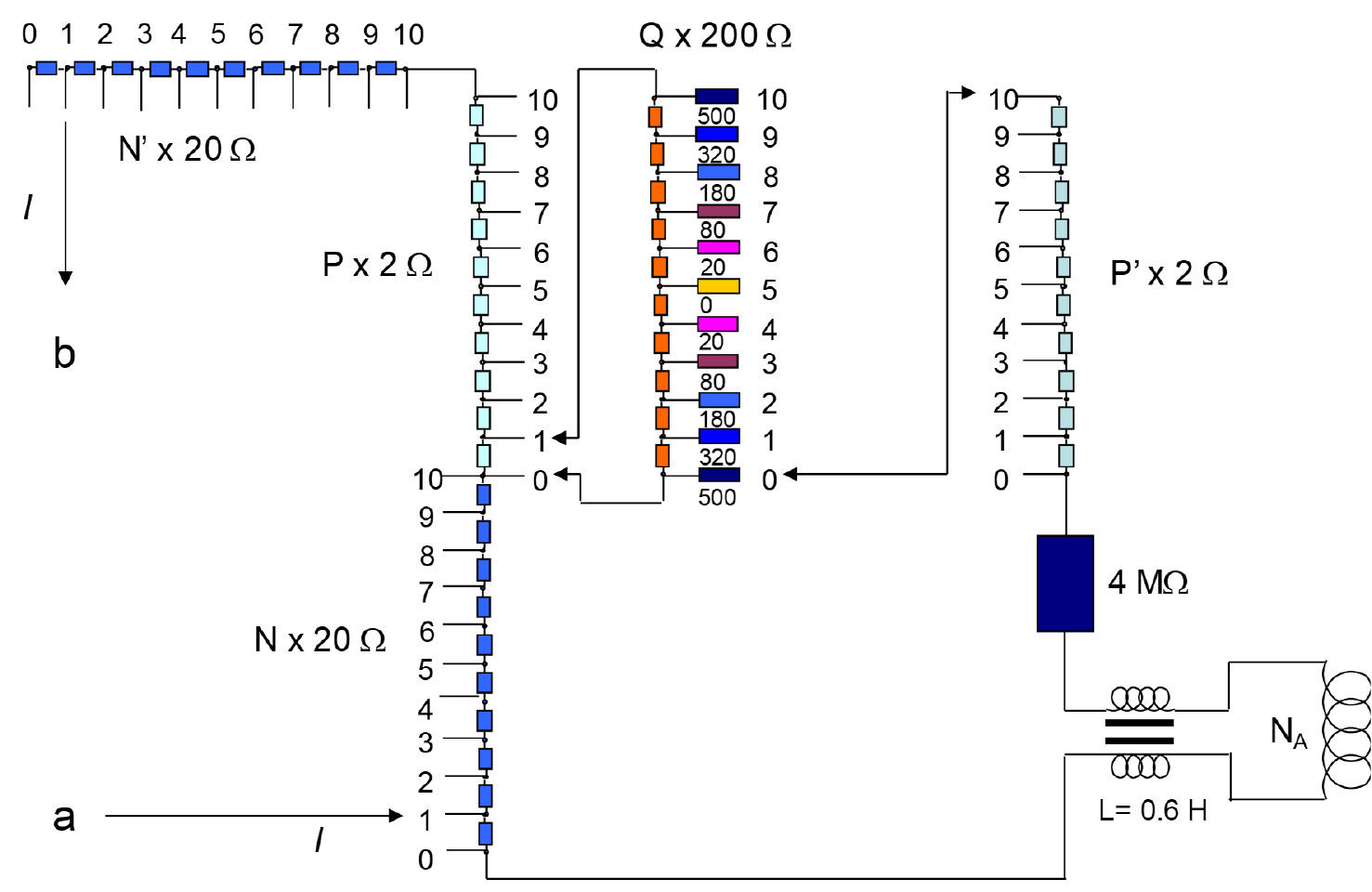}
\caption{Electrical scheme of the standard current divider (SCD). $N$, $P$, $Q$ are integers defining the setting of the SCD ($N'=10-N$, $P'=10-P$). A CMT is inserted between the SCD and the auxiliary winding to reduce the current noise circulation through ground. Connection points, a and b, are indicated in fig.\ref{fig1}.}\label{fig-SCD}
\end{figure}

Achieving a linear and stable SCD allowing the injection in the auxiliary winding of a fraction of the main current which ranges from 0 to $5\times10^{-5}$ with a minimal step of $5\times10^{-8}$, is a challenge. The SCD, described in fig.\ref{fig-SCD}, is made of three main series resistor networks ($10\times 20~\Omega$, $10\times 2~\Omega$, $10\times 200~\Omega$) and a large $\mathrm{4~M\Omega}$ division resistor. The nominal current fraction is given by: $\epsilon^\mathrm{nom}_{(N,P,Q)}= N\times5\times10^{-6}+P\times5\times10^{-7}+Q\times5\times10^{-8}$ where \emph{N}, \emph{P}, \emph{Q} are the integer values between 0 and 10 indexing the position of three selecting mechanical commutators (IEC MONACO commutators with gold-coated silver contacts).

The SCD is designed so that the selection of a given fraction does not require the disconnection of any of the resistors forming the three main networks: they remain soldered. Besides, the definition points of the fractions in a network, indexed by an integer, are physically realized by soldered wires. Moreover, high stability (drift lower than $10^{-5}$/year, in relative value), low temperature coefficient ($<0.6\times10^{-6}/^{\circ}$C) and hermetically-shielded Vishay resistors are used. The maximum power is dissipated in the 20 $\Omega$ resistors and remains below 2 mW. All these technical considerations ensure better stability and reproducibility of the SCD, even under load. The counterpart of the design is that non-linearities result from the variation of the \emph{P} and \emph{Q} parameters. Indeed, the triangle formed by resistors ($2~\Omega$, $Q\times 200~\Omega$, $(10-Q)\times 200~\Omega$) leads to the addition of a \emph{Q}-dependent resistance to the $\mathrm{4~M\Omega}$ resistance. This non-linearity is solved by adding a compensation resistance in series with the $\mathrm{4~M\Omega}$ resistance which is selected for each \emph{Q} value among the resistor network ($0~\Omega$, $20~\Omega$, $80~\Omega$, $180~\Omega$, $320~\Omega$, $500~\Omega$, $320~\Omega$, $180~\Omega$, $80~\Omega$, $20~\Omega$, $0~\Omega$). The division resistance defining the ratio also varies as a function of the \emph{P} parameter. This is compensated by selecting a fraction of the resistor network ($10\times 2~\Omega$) using the index \emph{P'}=10-\emph{P}. Finally, the fraction of the resistor network ($10\times 20~\Omega$) selected by \emph{N'} index adds to keep constant the total resistance of the SCD independently of the \emph{N} index change:  \emph{N'}=10-\emph{N}. This is useful to set the frequency bandwidth of the current source independently of the SCD setting. Let us note that a CMT is also introduced in the SCD circuit to reduce the current noise signals circulating from ground to the auxiliary winding (see picture in fig.\ref{FigAppendix-SCD} of Appendix \ref{annexe:SCD}).
\begin{figure}[h]
\centering
\includegraphics[width=3.0in]{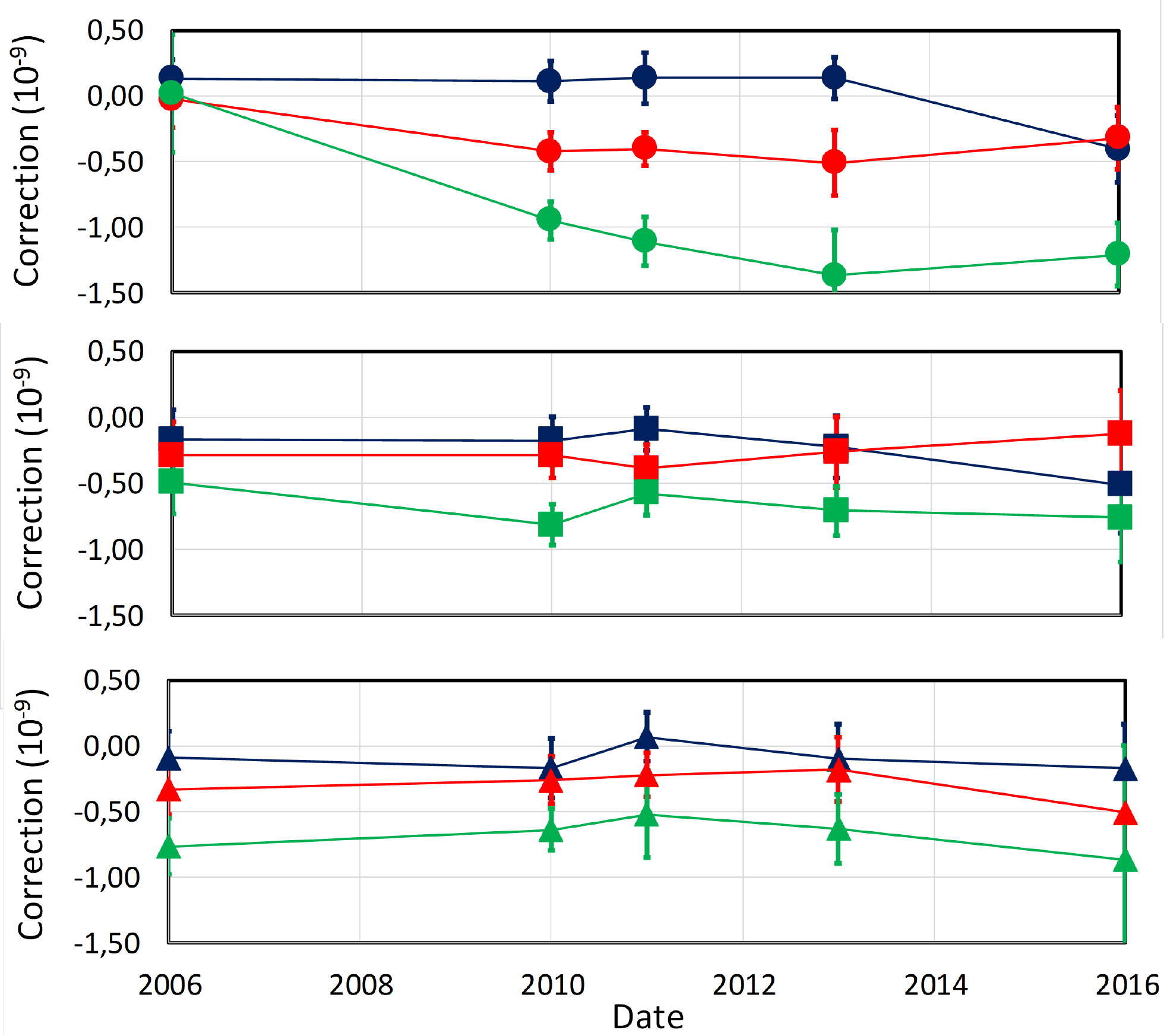}
\caption{Time evolution over ten years of the corrections ($\epsilon_{N,P,Q}-\epsilon^\mathrm{nom}_{N,P,Q})$, expressed in $10^{-9}$, that must be added to the nominal fractions, $(N\times5\times10^{-6})$ (circle), $P\times5\times10^{-7}$ (square) and $Q\times5\times10^{-8}$ (triangle) for three values of the integer index, 1 (blue), 5 (red), 9 (green). Error bars corresponds to standard uncertainties ($1\sigma$).}\label{fig-Corrections}
\end{figure}

The usual calibration of the SCD consists in measuring the three following sets of fractions: ($N$, $P=0$, $Q=0$), ($N=0$, $P$, $Q=0$) and ($N=0$, $P=0$, $Q$), with $N$, $P$, $Q$ varying from zero to ten. Any fraction is calibrated with a typical upper bound uncertainty of $3\times10^{-10}$ (see Appendix \ref{annexe:SCD}). Fig.\ref{fig-Corrections} reports on the time evolution over ten years of the determined corrections that must be added to the nominal values for the three ranges of fraction. Calibrations show that the fractions have remained close to their nominal value within $10^{-9}$, drifting each by no more than $5\times10^{-10}$, except for the highest one ($9\times5\times10^{-6}$) which has changed a bit more. The mean drift has been less than $1\times10^{-10}$/year between 2010 and 2016 (for future uses of the RB, the SCD will be calibrated every year). A fraction $\epsilon_{(N,P,Q)}$ is then obtained from four calibrated fractions by:
\begin{multline}
\epsilon_{(N,P,Q)}=\epsilon_{(N,P=0,Q=0)}+\epsilon_{(N=0,P,Q=0)}\\
+\epsilon_{(N=0,P=0,Q)}-2\epsilon_{(0,0,0)}.
\end{multline}
Its uncertainty, $u_\mathrm{B}^\mathrm{SCD}$, is about $0.5\times10^{-9}$. For more accurate resistance measurements, as those performed for international comparisons, the $\epsilon_{(N,P,Q)}$ fractions used can be specifically calibrated to within an uncertainty of $0.35\times10^{-9}$.
\subsection{The quadrature current divider}
The quadrature current divider (QCD), connected in series with the SCD (see fig.\ref{fig1}), injects an in-quadrature fraction $j\omega\epsilon^q=\alpha R_\mathrm{q}C_\mathrm{q}\omega$ of the main current in a CCC auxiliary winding of number of turns ${N_\mathrm{A}^q}$, as highlighted by equations \ref{Equation:AmpereTurnsQCD} and \ref{Equation:Phase}. More precisely, the main current is flowing through a $R_\mathrm{q}$ resistor. A $100~\Omega$ potentiometer connected in parallel allows the adjustment of the voltage fraction $\alpha$ biasing a $C_\mathrm{q}=235$ nF PTFE capacitor (its parallel resistance is higher than $2\times 10^{13}~\Omega$) in series with the CCC auxiliary winding. Depending on the amplitude of the quadrature compensation required, ${N_\mathrm{A}^q}$ is typically varied between 128 and 1600 and $R_\mathrm{q}$ is chosen equal to either $1~\Omega$ or $10~\Omega$ (exceptionally $100~\Omega$). For the measurement of the $100~\Omega/(R_\mathrm{K}/2)$ ratio, ${N_\mathrm{A}^q}=160$ and $R_\mathrm{q}=1~\Omega$. Prior to automatic measurements, the quadrature compensation is adjusted using the potentiometer so that the voltage overshoots observable during manual reversing of the current direction disappear. If the bridge is controlled by the ac signal of a lock-in oscillator, the QCD is adjusted to cancel the quadrature voltage signal.

In dc operation of the bridge, it is crucial that the QCD does not inject any current during data recording (or any in-phase current in ac operation of the bridge). Considering a current rise time of 0.5 s in dc operation, one can calculate that the injected current fraction drops down to no more than a few $10^{-14}$ ($10^{-13}$) of the main current after a waiting time of only 8 s for $R_\mathrm{q}=1~\Omega$ ($R_\mathrm{q}=10~\Omega$). For $N_\mathrm{A}=16$ and ${N_\mathrm{A}^q}=160$, the relative error on the resistance ratio caused by this residual current is ten times larger, however it remains negligible ($u_\mathrm{B}^\mathrm{QCD}<10^{12}$ for $R_\mathrm{q}=1~\Omega$). Let us note that the quadrature current divider is also equipped with a CMT to reduce current noise circulation (see fig.\ref{FigAppendix-SCD} of Appendix section \ref{annexe:SCD}).
\section{Shielding and guarding}
\begin{figure}[h!]
\centering
\includegraphics[width=3.0in]{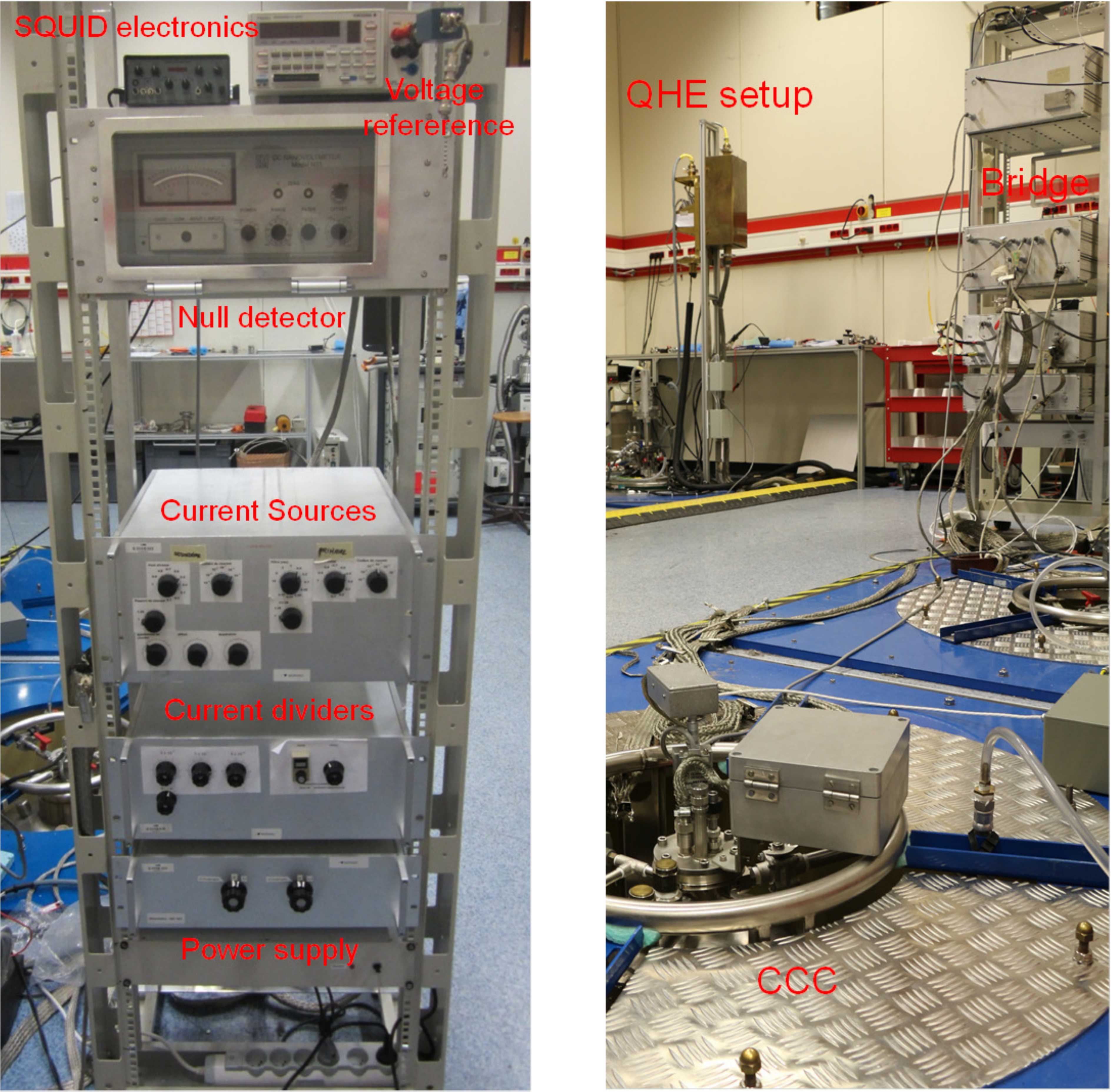}
\caption{Left: Picture of the resistance bridge. Right: Picture of the CCC winding switching box at the top of the cryostat.}\label{Fig-FullBridge}
\end{figure}
To shield against noise, the sensitive elements of the RB (the null detector, the current sources, the current dividers, the power supply) were integrated in metallic boxes (see fig. \ref{Fig-FullBridge}). Both with the QHR and the CCC setups, they are connected at ground which is materialized by the copper floor of the laboratory Faraday cage. The continuity of the shielding between the different elements is ensured by the connection cables, the metallic sheath of which is also connected at ground as schematized in fig.\ref{fig1}. Cables are connected using shielded PTFE-insulated Fischer$^{\scriptsize\textregistered}$ connectors. This directs any leakage current between wires at different potentials towards the ground. In normal operation, the ground is connected both to the low potential of the resistance $R_\mathrm{S}$ (position A in fig.\ref{fig1}) and to the case of the EM detector (there is no common mode voltage). The leakage current, $I_g$ therefore short-circuits the lowest resistance (black arrow) which reduces its impact. This grounding is usually efficient to limit the leakage current contribution to the type B relative uncertainty below $10^{-9}$ for the measurement of the $100~\Omega/(R_\mathrm{K}/2)$ ratio. The ground can also be connected to the low potential of the secondary winding (position B in fig.\ref{fig1}). The CMT remains efficient and the SQUID operation stays optimal. In this case, the leakage current short-circuits both the resistor and the winding of the secondary circuit (grey arrow) and therefore does not degrade the measurement accuracy.

Let us remark that the QCD adjustment depends on the ground point position since it changes the quadrature leakage current. Settings $R_\mathrm{q}=1~\Omega$ and ${N_\mathrm{A}^q}=128,160$ are generally adapted for ground in position B since capacitance leakage current are fully deviated. As expected, larger values of $R_\mathrm{q}$ or ${N_\mathrm{A}^q}$ are required for ground in position A and $R_\mathrm{S}\geq 1~k\Omega$.
%But, this is at the expense of a common mode voltage, although weak, existing between the ground and the low potential of the null detector due to the small resistance (about 2 $\Omega$) of the winding. To minimize its effect on the measurement accuracy, the case and the low potential of the null detector are short-circuited. Besides, capacitance hand-effects are cancelled because the null detector is itself placed in a grounded metallic box.
\section{Resistance ratio Measurements}
\subsection{Noise spectrum and SQUID feedback stability}
\label{section:NoiseSpectrum}
The operation stability of the resistance bridge, i.e. of the SQUID, was demonstrated for the measurement of many resistance ratios. Table \ref{tableau:Settings} presents the settings of the RB which were used for the measurements reported in this paper.
\begin{table}[h]
\newcolumntype{M}[1]{>{\centering\arraybackslash}m{#1}}
\begin{center}
\begin{tabular}{|c|c|c|c|c|}
  \hline
  \textbf{Ratio}&$\bm{N_\mathrm{P}}$&$\bm{N_\mathrm{S}}$&$\bm{N_\mathrm{A}}$&\textbf{SCD position}\tabularnewline \hline
  \textbf{100 $\Omega$/($R_\mathrm{K}/2$)}&2065&16&16&secondary source\tabularnewline \hline
  \textbf{200 $\Omega$/($R_\mathrm{K}/2$)}&2065&32&16&secondary source\tabularnewline \hline
  \textbf{100 $\Omega$/10 k$\Omega$}&1600&16&16&secondary source\tabularnewline \hline
  \textbf{1 k$\Omega$/($R_\mathrm{K}/2$)}&2065&160&160&secondary source\tabularnewline \hline
  \textbf{100 $\Omega$/1 k$\Omega$}&160&16&16&secondary source\tabularnewline \hline
  \textbf{1 k$\Omega$/1 k$\Omega$}&160&160&64&secondary source\tabularnewline \hline
  \textbf{10 k$\Omega$/1 M$\Omega$}&1600&16&16&secondary source\tabularnewline \hline
  \textbf{1 $\Omega$/100 $\Omega$}&1600&16&1600&primary source\\ \hline
\end{tabular}
\end{center}
\caption{Settings of the RB for several resistance ratio measurements.}
\label{tableau:Settings}
\end{table}
\begin{figure}[h]
\centering
\includegraphics[width=3.5in]{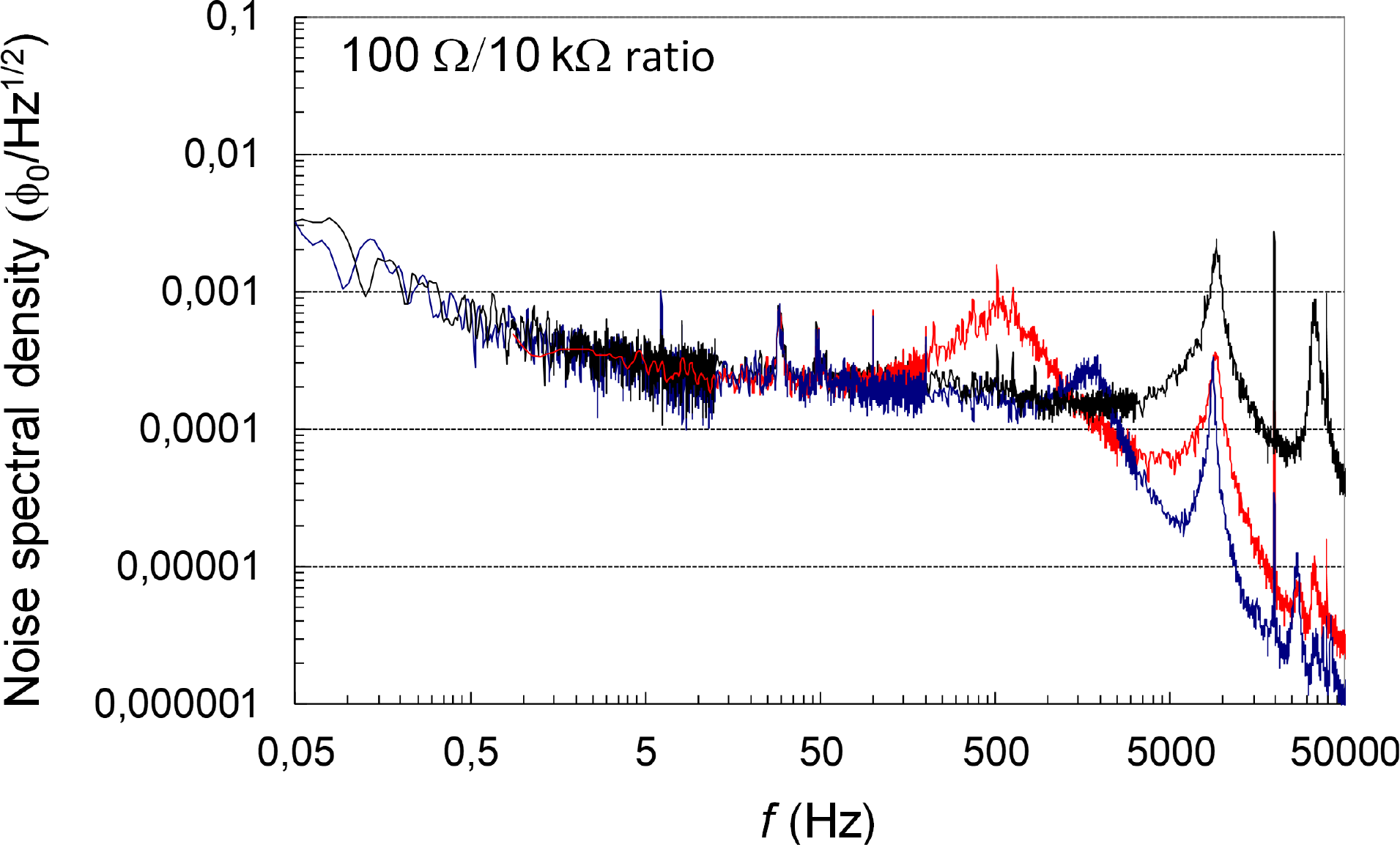}
\caption{Noise spectral density measured by the SQUID \emph{versus} frequency \emph{f} for the measurement of the 100 $\Omega$/10 k$\Omega$ ratio. SQUID operating in internal feedback mode 5 (black), in external feedback mode 5s (red), and in external feedback mode 500 (blue).}\label{Fig-NoiseCCC10k100}
\end{figure}

Here, we focus on the analysis of the noise spectral density, expressed in $\mathrm{\phi_0/Hz^{1/2}}$, determined by the Quantum Design SQUID operating in different internal and external feedback modes for the measurement of the 10 k$\Omega$/1 M$\Omega$, 100 $\Omega$/10 k$\Omega$ and 1 $\Omega$/100 $\Omega$ ratios (the quadrature divider was not used for these tests).

Fig.\ref{Fig-NoiseCCC10k100} shows the noise spectral density obtained for the measurement of the 100 $\Omega$/10 k$\Omega$ ratio using current ranges $10$ mA/100 $\mu$A. In closed feedback mode operation, the measured noise corresponds to the uncorrelated noise contributions of both current sources since the current ratio $I_\mathrm{S}/I_\mathrm{P}$ is adjusted to within $2\times10^{-6}$ to cancel the ampere.turn unbalance of the CCC, i.e. the magnetic flux in the SQUID. Let us note that the residual magnetic flux noise crossing the SQUID itself has a much lower level because of the real-time compensation by the feedback signal. It is given by the combination of the intrinsic SQUID noise (3 $\mathrm{\mu\phi_0/Hz^{1/2}}$), the environmental noise directly captured by the SQUID and the current source noise divided by the open-loop amplification gain. This latter contribution is negligible. The two others, which manifest in the CCC alone and disconnected, give a contribution of about 10 $\mathrm{\mu\phi_0/Hz^{1/2}}$ as observed in fig.\ref{Fig-NoiseCCCAlone}.

In internal feedback mode 5, the SQUID open-loop bandwidth of 50 kHz allows measuring the noise level up to the frequency resonances of the CCC. The noise amplitude is above a noise floor level of about 140 $\mathrm{\mu\phi_0/Hz^{1/2}}$. This bottom level is notably explained by the Johnson-Nyquist noise, of 120 $\mathrm{\mu\phi_0/Hz^{1/2}}$, generated by the $R_\mathrm{C}=50$ k$\Omega$ resistor defining the 100 $\mu$A current range of the primary current source. Below 10 Hz, the noise increase is mainly caused by the $1/f$ voltage noise of the operational amplifiers (OPA111BM) which drives the $R_\mathrm{C}$ resistor. The operation in external feedback mode 5s is very stable. As can be observed, the noise spectrum is similar but is characterized by a lower frequency bandwidth limited to about 500 Hz which manifests itself by a small peak above which the signal then decreases of 20 dB by decade. This can be explained by the reduction to 500 Hz of the SQUID open-loop bandwidth, which is recommended to improve the stability in presence of short electro-magnetic transients, and the 1 kHz frequency bandwidth of the feedback circuit (see section \ref{Subsection:Feedback}). The operation in external feedback mode 500 also turns out to be stable. It is characterized by a noise spectrum having a higher cutoff frequency of about 1 kHz determined by the feedback circuit only (the SQUID open-loop bandwidth is of 50 kHz). In some measurement configurations, this larger frequency bandwidth can improve the stability of the bridge operation against acoustic noises. But generally, the bridge is operated using the external feedback mode 5s.
 \begin{figure}[h]
\centering
\includegraphics[width=3.5in]{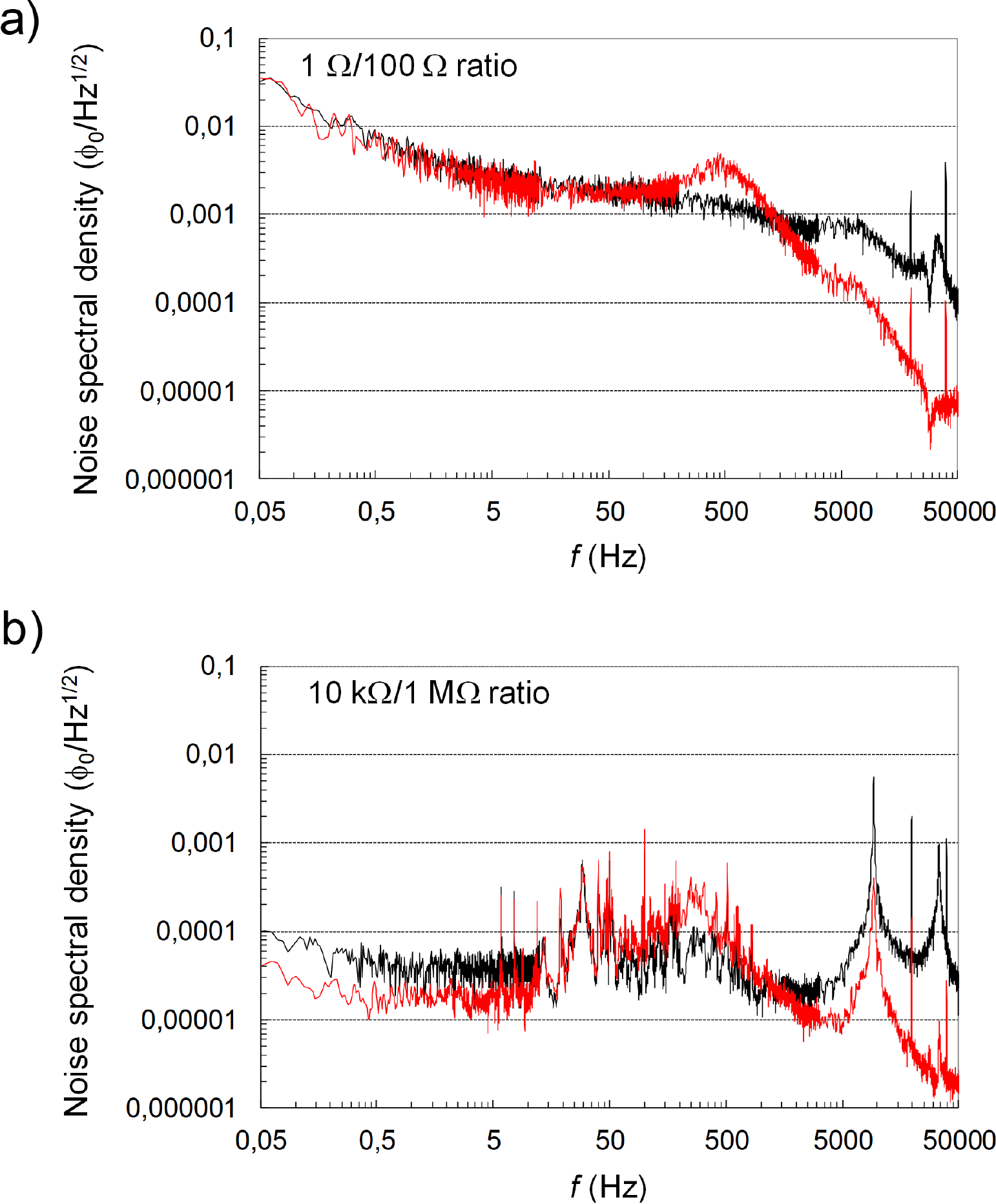}
\caption{Noise spectral density measured by the SQUID \emph{versus} frequency \emph{f} for the measurement of the 1 $\Omega$/100 $\Omega$ ratio a) and 10 k$\Omega$/1 M$\Omega$ ratio b). SQUID operating in internal feedback mode 5 (black) and in external feedback mode 5s (red).}\label{Fig-NoiseCCCAutres}
\end{figure}

Fig.\ref{Fig-NoiseCCCAutres}a and b demonstrate stability of operation of the external SQUID feedback in the measurements of ratios 1$\Omega$/100 $\Omega$ and 10 k$\Omega$/1 M$\Omega$ respectively. The base noise level is larger for the measurement of the 1 $\Omega$/100 $\Omega$ ratio (above $\mathrm{{2~m\phi_0/Hz^{1/2}}}$). This comes from the reduction to 5 k$\Omega$ of the resistor $R_\mathrm{C}$ defining the 1 mA range used to supply the 100 $\Omega$ resistor. Conversely, the current sources are feebly noisy in the measurement configuration of the 10 k$\Omega$/1 M$\Omega$ ratio because of the $R_\mathrm{C}=5$ M$\Omega$ resistor defining the 1 $\mu$A range. One can observe in fig.\ref{Fig-NoiseCCCAutres}b, a white noise level of no more than $\mathrm{20~\mu\phi_0/Hz^{1/2}}$ between 0.2 Hz and 6 Hz. This low base noise level allows observing the manifestation of moderate mechanical resonances in the range from 10 Hz and 1 kHz.
\subsection{Measurement protocol and type A uncertainty}
\begin{figure}[h]
\centering
\includegraphics[width=3.5in]{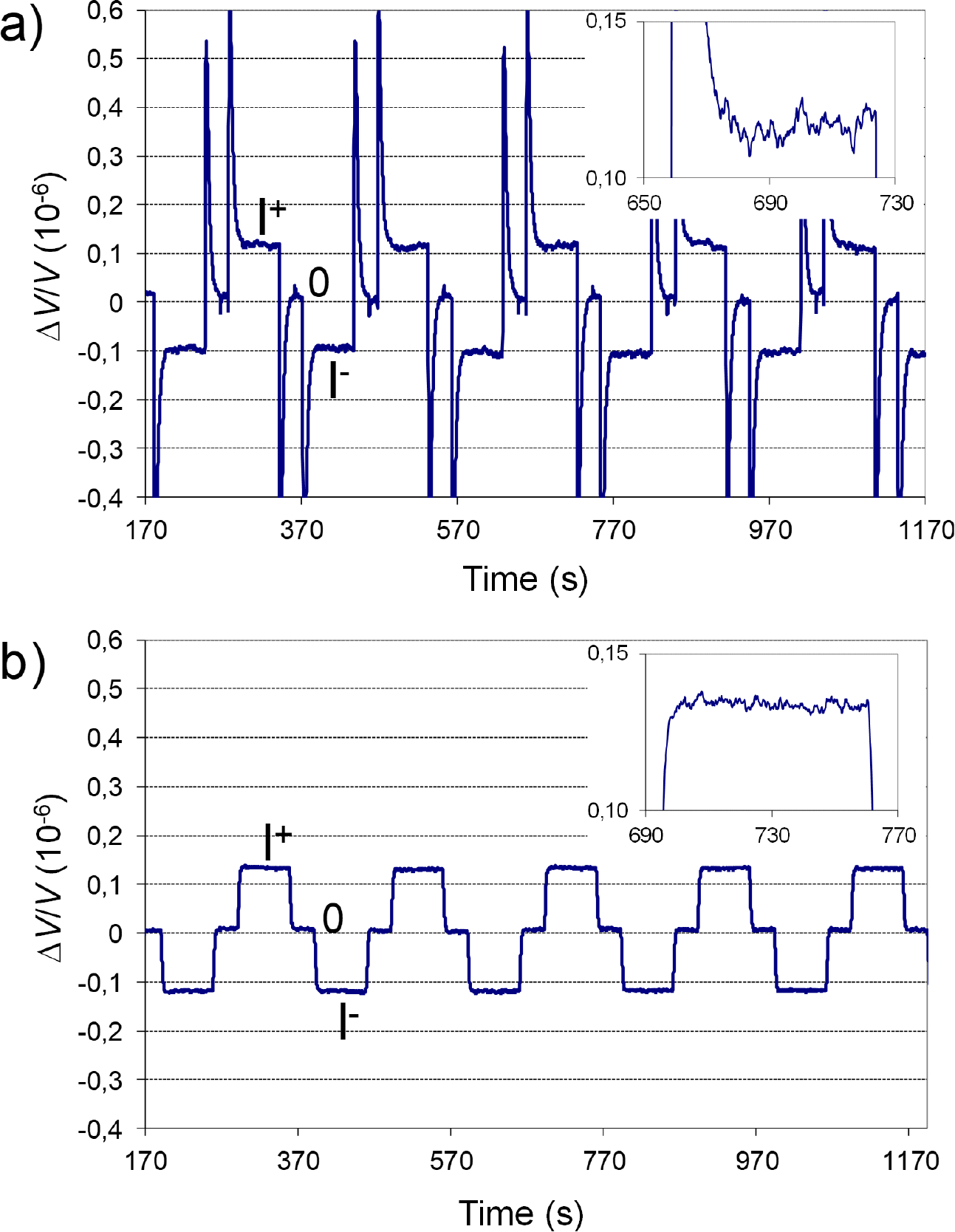}
\caption{Measurement of the resistance ratio $r_R=100~\Omega/(R_\mathrm{K}/2)$ using the old a) and the new b) LNE bridge: relative voltage deviation $\Delta V/V$ as a function of time for several ($\mathrm{I^+}$, 0, $\mathrm{I^-}$) sequences. The signal period is about 200 s. Insets: enlargment of voltage plateaus. The following settings were used: $N_\mathrm{P}=1936$, $N_\mathrm{S}=15$ and $N_\mathrm{A}=15$ for the older bridge and $N_\mathrm{P}=2065$, $N_\mathrm{S}=16$, $N_\mathrm{A}=16$ and $N_\mathrm{A}^q=1600$ for the new bridge. $\epsilon$ fractions are chosen to obtain similar deviation amplitude, $\Delta V/V$, for both bridges. $\epsilon^q$ setting of the new bridge QCD is optimized to cancel overshoots occurring during current reversals.}\label{Fig-StabilitySteps}
\end{figure}
Measurements of the resistance ratio $r_R=100~\Omega/(R_\mathrm{K}/2)$ were performed using the old and the new LNE bridges. The primary current circulating through the GaAs/AlGaAs-based quantum resistance standard is set to $I_\mathrm{P}=70~\mu$A. For a $\epsilon$ fraction of the SCD which differs from $\epsilon_\mathrm{eq}$ (giving $\Delta V=0$), a finite voltage $\Delta V$ can be detected by the null detector. For all resistance comparisons reported in this paper, the voltage, $\Delta V$, is measured by a EMN 11 nanovoltmeter (time constant 1.3 s, 3 $\mu$V range), the isolated output of which is recorded by a $6^{1/2}$ digit Keithley 2000 multimeter (sample rate of 4 Hz). The relative voltage $\Delta V/V$, where $V=R_\mathrm{P}I_\mathrm{P}$, is related, at the first order, to the deviation $(\epsilon-\epsilon_\mathrm{eq})$ by:
\begin{equation}
\Delta V/V=(\epsilon-\epsilon_\mathrm{eq})\frac{N_\mathrm{A}}{N_\mathrm{S}}.
\end{equation}
Let us note that, reversely, $\Delta V/V$ could be interpreted as a relative deviation of the ratio $r_R$ to the value $r_{R_\mathrm{eq}}$ giving $\Delta V=0$ for the fraction $\epsilon$. $\Delta V/V$ measured as a function of time during several ($\mathrm{I^+}$, 0, $\mathrm{I^-}$) sequences is reported in Fig.\ref{Fig-StabilitySteps}a) and b) for the old and the new LNE bridge respectively. The signal period, of about 200 s, is imposed by the low-speed capability of the older bridge. The comparison of both data first shows the lower noise level and better stability achieved in measurements performed with the new bridge. This comes not only from the better performance of the current source electronics but also from the lower noise level of the CCC. Second, it demonstrates the efficiency of the QCD in cancelling any voltage overshoot during the current switchings. This allows increasing the current reversal frequency in order to reduce the impact of the voltage offset drift and of the 1/$f$ SQUID noise.
\begin{figure}[h]
\centering
\includegraphics[width=3.5in]{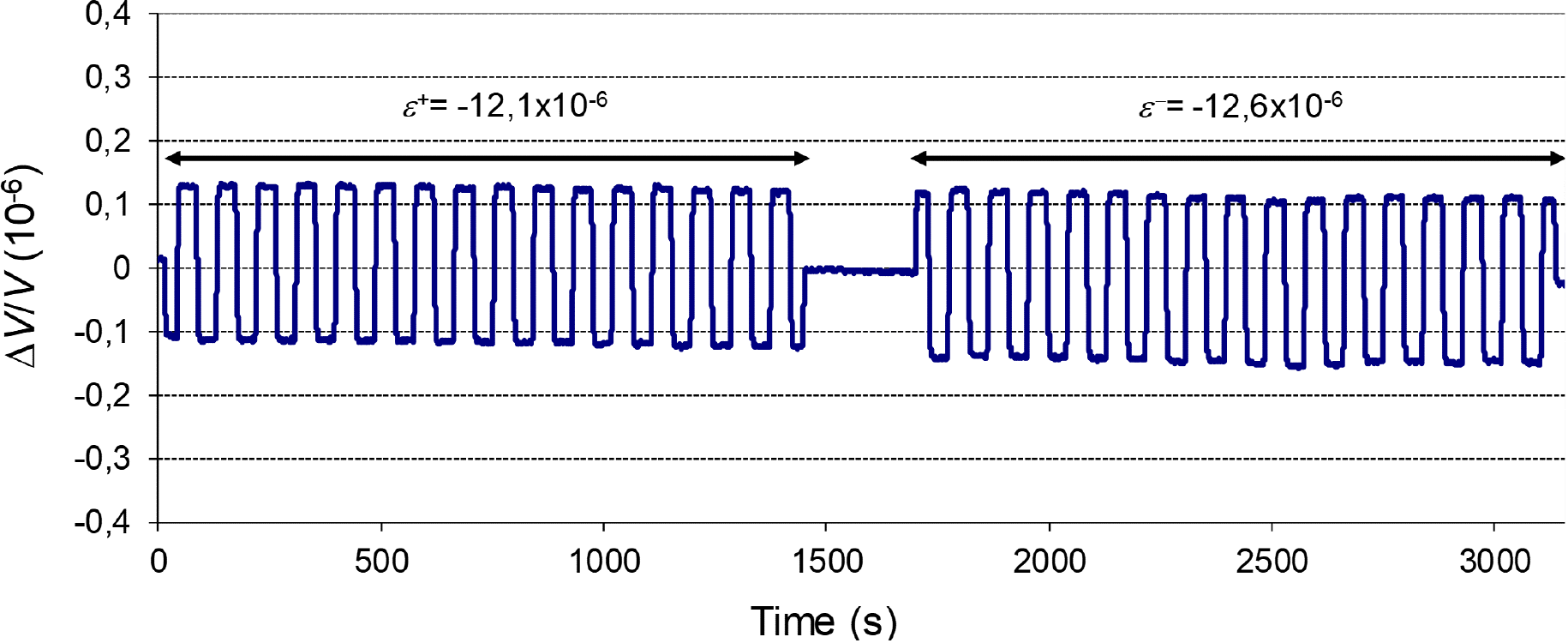}
\caption{Measurement of the resistance ratio $r_R=100~\Omega/(R_\mathrm{K}/2)$ using the new LNE bridge with a primary current $I_\mathrm{P}=70~\mu$A: relative deviation $\Delta V/V$ as a function of time for 16 ($\mathrm{I^+}$, $\mathrm{I^-}$, $\mathrm{I^+}$) sequences and two successive settings, $\epsilon^+=-12.1\times10^{-6}$ and $\epsilon^-=-12.6\times10^{-6}$, of the SCD fractions.}\label{Fig-QuickSequence}
\end{figure}

Fig.\ref{Fig-QuickSequence} shows the typical data record adopted for the measurement of the resistance ratio $r_R=100~\Omega/(R_\mathrm{K}/2)$ with the new bridge. It consists of two successive acquisitions of voltage measurements that are obtained for the two settings of the SCD fractions, $\epsilon^+=-12.1\times10^{-6}$ and $\epsilon^-=-12.6\times10^{-6}$ respectively. Each acquisition is made of 16 ($\mathrm{I^+}$, $\mathrm{I^-}$, $\mathrm{I^+}$) sequences of current reversal that are used to remove voltage offsets. The current rise time and the waiting time before acquisition are set to 0.5 s and 12 s respectively. A mean voltage value is calculated from the average of the 16 values $[V(I^+)_1+V(I^+)_3-2V(I^-)_2]/4$, where j=1,2,3 indexes the current state of each sequence and $V(I^\pm)_j$ is itself the mean value of 60 voltage measurements. The resistance ratio is then obtained from the $\epsilon_\mathrm{eq}$ value calculated from the two mean voltages, $V^{\epsilon^-}$ and $V^{\epsilon^+}$, using the relation:
\begin{equation}
\epsilon_\mathrm{eq}=\epsilon^- + (\epsilon^+-\epsilon^-)\times\frac{|V^{\epsilon^-}|}{(|V^{\epsilon^-}|+|V^{\epsilon^+}|)}.
\end{equation}

Owing to the new bridge performances, the period of the signal was therefore reduced to about 70 s and the zero crossing step was removed. The result is that the ratio between the acquisition time and the total experiment time is increased from about 50 percents with the older bridge to 75 percents with the new bridge, which is favorable to a reduction of the type A uncertainty. Similar operation of the new resistance bridge is achieved for the measurement of other ratios using the settings reported in table \ref{tableau:Settings} and adapted QCD adjustments: 200 $\Omega$/$(R_\mathrm{K}/2)$, 100 $\Omega$/10 k$\Omega$, 1 k$\Omega$/1 k$\Omega$, 1 k$\Omega$/$(R_\mathrm{K}/2)$ and 100 $\Omega$/1 k$\Omega$.
\begin{figure}[h]
\centering
\includegraphics[width=3.5in]{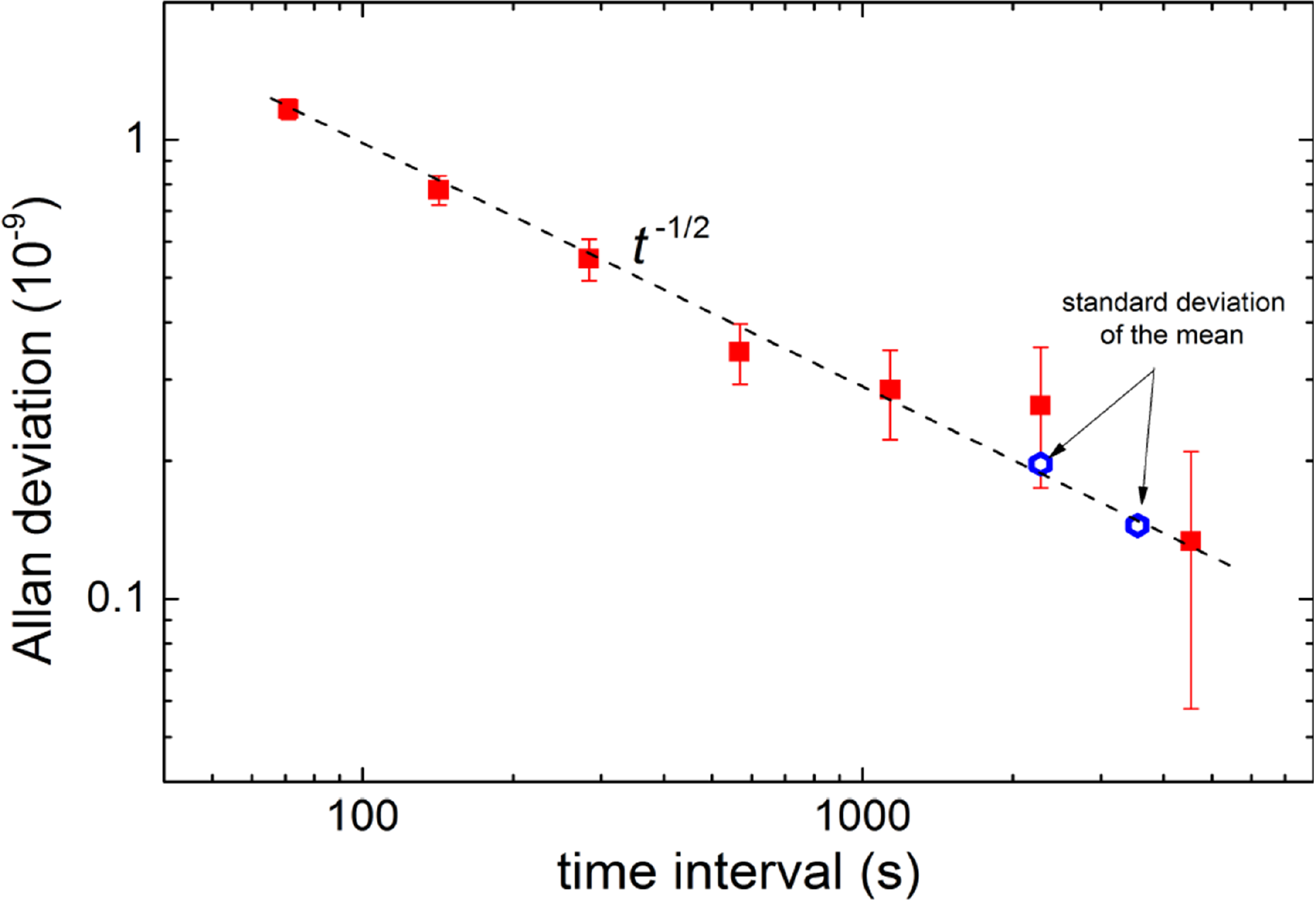}
\caption{Measurement of the resistance ratio $r_R=100~\Omega/(R_\mathrm{K}/2)$ using the new LNE bridge with a primary current $I_\mathrm{P}=50~\mu$A and a signal period of 70 s. From an experiment described in supplementary of \cite{Ribeiro2015}. Standard Allan deviation of $r_R$, expressed in relative value, as a function of the acquisition time $t$ (red square). $t^{-1/2}$ adjustment of data (black dashed line). Standard deviation of the mean (blue open hexagon).}\label{Fig-Allan}
\end{figure}

The noise performance of the new LNE bridge is demonstrated by the calculation of the Allan standard deviation\cite{Allan1987,Witt2005} of $r_R=100~\Omega/(R_\mathrm{K}/2)$ from the statistical analysis of the voltage measurements performed using a primary current $I_\mathrm{P}=50~\mu$A. The evolution of this quantity, expressed in relative value, is reported in fig.\ref{Fig-Allan} as a function of the experience time $t$. It follows a $t^{-1/2}$ law over measuring times longer than one hour. This shows that the white noise is dominant and that the standard deviation of the mean can be used as an estimate of the type A relative uncertainty, $u_\mathrm{A}$. It follows that $u_\mathrm{A}=0.15\times10^{-9}$ can be achieved for the measurement of the ratio $r_R=100~\Omega/(R_\mathrm{K}/2)$ using a current of $50~\mu$A after an observation time of one hour. This is five times less than the best uncertainty achievable with the older bridge. This improvement relies not only on the quicker measurement protocol but also on the lower noise of the current source electronics and of the CCC. Let us remark that the contribution of the CCC to the voltage noise at the terminals of the null detector is no more than $\mathrm{0.5~nV/Hz^{1/2}}$. This is more than ten times lower than the EMN 11 nanovoltmeter contribution, of about $\mathrm{7~nV/Hz^{1/2}}$, which clearly limits the bridge type A uncertainty. Considering a ratio of 75 percents between the acquisition time and the total experiment time, its noise leads to a calculated type A relative uncertainty of $1.15\times10^{-8}t^{-1/2}$ in good agreement with the Allan deviation reported in fig.\ref{Fig-Allan}.
\subsection{Uncertainty budget}
Table \ref{tableau:Uncertainty} itemizes the uncertainty budget established for the measurement of the ratio $100~\Omega/(R_\mathrm{K}/2)$. The Type A contribution and Type B contributions related to CCC accuracy, SCD calibration and QCD have already been discussed. Below are considered other Type B contributions including those caused by SQUID feedback and leakage current.
\begin{table}[h]
\newcolumntype{M}[1]{>{\centering\arraybackslash}m{#1}}
\begin{center}
\begin{tabular}{|c|c|c|}
  \hline
  \textbf{Contributions}&\textbf{Name}&\textbf{Contributions ($10^{-9}$, k=1)} \tabularnewline \hline
  \textbf{Type A (1 hour)}&$\bm{u_\mathrm{A}}$&\textbf{0.15}\tabularnewline \hline
  \textbf{Type B}&$\bm{u_\mathrm{B}}$& \textbf{0.7 (A), 0.5 (B)}\tabularnewline \hline
  CCC accuracy&$u_\mathrm{B}^\mathrm{CCC}$& $<0.1$\tabularnewline \hline
  SQUID feedback accuracy&$u_\mathrm{B}^\mathrm{FB}$& $<0.01$\tabularnewline \hline
  SCD calibration&$u_\mathrm{B}^\mathrm{SCD}$& $<0.5$\tabularnewline \hline
  QCD &$u_\mathrm{B}^\mathrm{QCD}$& $<0.001$\tabularnewline \hline
  Leakage to ground&$u_\mathrm{B}^\mathrm{g}$&$\sim 0.5$ (A), 0.1 (B)\tabularnewline \hline
  \textbf{Combined Uncertainty}&$\bm{u_\mathrm{C}}$&\textbf{0.7 (A), 0.6 (B)}\\ \hline
\end{tabular}
\end{center}
\caption{Preliminary relative uncertainty budget for the $100~\Omega/(R_\mathrm{K}/2)$ ratio considering a $I_\mathrm{P}=50~\mu$A measurement current.}
\label{tableau:Uncertainty}
\end{table}
\subsubsection{SQUID feedback accuracy}
The accuracy of the resistance ratio measurement depends also on the SQUID feedback accuracy in cancelling the total ampere.turns number, i.e. in setting the current ratio $r_I=I_\mathrm{S}/I_\mathrm{P}$ to the target ratio given by the equation \ref{Equation:AmpereTurns}. A setpoint error comes from the finite value of the open loop gain $G_\mathrm{OLG}$ of the SQUID electronics and the imperfect adjustment of the current ratio prior to the SQUID feedback operation. A type B relative uncertainty, $u_B^{FB}$, lower than $10^{-11}$ is determined for usual prior adjustment level of the current ratio (see Appendix \ref{annexe:CLG}).
\subsubsection{Leakage current to ground}
Several experiments were carried out to estimate the impact of leakage current to ground on the measurement accuracy. No leakage current could be unveiled by comparing measurements of the resistance ratios $100~\Omega/(R_\mathrm{K}/2)$ and $200~\Omega/(R_\mathrm{K}/2)$, performed with the ground either connected in position A ($I_g$ parallel to $R_\mathrm{S}$) or position B ($I_g$ fully deviated) since no significant deviation was found within a relative uncertainty of about $0.8\times10^{-9}$. Comparisons were therefore repeated with a larger secondary resistance $R_\mathrm{S}$=1 k$\Omega$ to amplify the effect of leakage currents and make it detectable. From the comparison of the two measurements of the ratio 1 k$\Omega/$1 k$\Omega$ performed with the resistors interchanged, a relative deviation of $(-3\pm 0.4)\times10^{-9}$ is found for the measurement of one resistance ratio for the ground in position A which reduces to $(-0.04\pm 0.3)\times10^{-9}$ for the ground in position B. A similar relative deviation of a few $10^{-9}$ is also found by comparing the measurements of the ratio 1 k$\Omega/(R_\mathrm{K}/2)$ obtained for both ground positions. One concludes that a significant negative discrepancy of a few $10^{-9}$ resulting from leakage currents exists but can nevertheless be fully cancelled by moving the ground in position B. From these comparisons, a Type B relative uncertainty, $u_B^{g}$, of about $0.5\times10^{-9}$, i.e. ten times lower, can therefore be deduced for the measurement of the 100 $\Omega/(R_\mathrm{K}/2)$ ratio with the ground in position A. It falls below $0.1\times10^{-9}$ by connecting the ground in position B. Let us remark that further characterizations are required to explain the origin of the larger than expected leakage current to ground. Indeed, its magnitude cannot be directly explained from the value of the leakage resistance to ground of the whole measurement system (bridge, CCC, resistance standards), which is measured to be higher than 4 T$\Omega$. Moreover, the leakage resistance to ground of the current sources alone is higher than 80 T$\Omega$ confirming the high galvanic insulation of the operational amplifiers.
\subsubsection{Small contributions}
No significant effect of the current reversal duration (from $I^+$ to $I^-$ and reversely) was found within a relative uncertainty of $0.35\times10^{-9}$ by varying its value from 12 s to 24 s while keeping the same acquisition time. The absence of observable asymmetry of the voltage deviations measured by the null detector for $\mathrm{I^+}$ and $\mathrm{I^-}$ current directions indicates small effect of noise rectification in the measurements discussed. However, further work is required to generalize this conclusion to any ratio measurement.\\
\\
To conclude, the total type B relative uncertainty is estimated to be either $0.7\times10^{-9}$ or $0.5\times10^{-9}$ depending on whether the ground is connected to position A or B. The type A uncertainty being much smaller, the standard combined relative uncertainty is significantly below $10^{-9}$. Further reduction of the measurement uncertainty will come from the improvement of the current divider calibration.
\subsection{Validation of measurement performance}
Table \ref{tableau:Comparisons} reports on the deviations between the measurements of the ratios $100~\Omega/(R_\mathrm{K}/2)$ and $100~\Omega/10~k\Omega$ performed using the new and the old bridges. It shows that there is no significant discrepancy within a combined relative uncertainty below $1.5\times10^{-9}$. Let us note that the comparison uncertainty is limited by the larger type A uncertainty of the older bridge. This agreement between the two resistance bridges, which differ not only by their electronics but also by their CCC and standard current divider, make us very confident in our measurements of these resistance ratios. It consolidates the uncertainty budget described previously in table \ref{tableau:Uncertainty}.
\begin{table}[h]
\newcolumntype{M}[1]{>{\centering\arraybackslash}m{#1}}
\begin{center}
\begin{tabular}{|c|c|c|}
  \hline
  \textbf{Ratio}&\textbf{$100~\Omega/(R_\mathrm{K}/2)$}&\textbf{$100~\Omega/10~k\Omega$} \tabularnewline \hline
  \textbf{Relative deviation}&$(-1.4\pm1.5)\times10^{-9}$&$(-0.7\pm1.4)\times10^{-9}$\\ \hline
\end{tabular}
\end{center}
\caption{Relative deviations with combined standard uncertainties (k=1) between the measurements of the resistance ratio performed by the new and the older bridges.}
\label{tableau:Comparisons}
\end{table}

Moreover, the measured value of the ratio $100~\Omega/(R_\mathrm{K}/2)$ and the ratio value deduced by combining the measurements of the 1 k$\Omega/(R_\mathrm{K}/2)$ and 100 $\Omega$/1 k$\Omega$ ratios are found in agreement within a relative uncertainty of $1.2\times10^{-9}$ (with the ground in position B). This consistency check confirms the small impact of the leakage current in position B of the ground and that the $10^{-9}$ relative uncertainty also applies to the measurement of other resistance ratios than $100~\Omega/(R_\mathrm{K}/2)$.

Besides, the new LNE bridge was used to perform accurate universality tests of the QHE\cite{Lafont2015,Ribeiro2015}. The agreement of the quantized Hall resistance, $R_\mathrm{H}$, measured in GaAs and graphene devices was demonstrated with a record\cite{Ribeiro2015} relative uncertainty of $8\times10^{-11}$. This result was obtained by comparing the two measurements of the ratio $100~\Omega/(R_\mathrm{H}/2)$ carried out using a 100 $\Omega$ transfer resistor. This performance therefore emphasizes the low-noise level and the reproducibility of the measurement bridge, rather than its accuracy since many Type B uncertainty contributions are cancelled by the design of the comparison protocol. The capability of the resistance bridge to perform measurements at low frequency (2 Hz) allowed the determination of the temperature evolution of the quantized Hall resistance in graphene during dynamic temperature drift\cite{Lafont2015}. Finally, many elements of the resistance bridge, i.e. the CCC, the current source and the current divider, were also used to build the programmable quantum current generator that allowed a practical realization of the ampere from the elementary charge with a $10^{-8}$ relative uncertainty\cite{Brun-Picard2016}.
\section{Conclusion}
A new comparison resistance bridge based on a CCC was built at LNE. It is based on low-noise synchronized current sources, a new CCC with a very low-noise level of $\mathrm{80~pA.t/Hz^{1/2}}$, a standard current divider characterized by a one-year stability of the fractions within the calibration uncertainty of $0.3\times10^{-9}$ and also a quadrature current divider which cancels voltage overshoots during current transitions. Accurate measurements of the resistance ratios, $100~\Omega/(R_\mathrm{K}/2)$, $\mathrm{100~\Omega/10~k\Omega}$, $200~\Omega/(R_\mathrm{K}/2)$, 1 k$\Omega$/1 k$\Omega$, 1 k$\Omega$/$(R_\mathrm{K}/2)$ and 100 $\Omega$/1 k$\Omega$ were achieved. Besides, stable operation of the resistance bridge was also demonstrated for the measurement not only of the $\mathrm{10~k\Omega/1~M\Omega}$ ratio but also of the $1~\Omega/100~\Omega$ ratio (with current dividers inserted in the primary circuit). Next work will report on measurement accuracy of these two resistance ratios.

The $100~\Omega/(R_\mathrm{K}/2)$ ratio can be measured with a type A relative uncertainty below $0.2\times 10^{-9}$ within one hour measurement time. This performance results not only from the lower noise of the bridge but also from the optimization of the data acquisition thanks to the quadrature current divider. Further improvement would require a lower-noise null detector than the EMN 11. The total type B relative uncertainty, estimated considering main contributions, is of $0.5\times10^{-9}$. This leads to a standard combined relative uncertainty of only $0.6\times10^{-9}$. A slightly smaller combined standard uncertainty is expected at term by implementing a new calibration method of the standard current divider.
\appendices
\section{CCC calculations}
\label{annexe:CCC}
The sensitivity of the CCC, $S_\mathrm{CCC}$, depends on the number of turns $N_\mathrm{PC}$ of the pick-up coil coupled to the CCC through the relationship:\\
\begin{equation}
S_\mathrm{CCC}=(2/k)N_\mathrm{PC}S_\mathrm{SQ},
\end{equation}
where $k$ is the coupling constant between the CCC and the pickup coil, and $S_\mathrm{SQ}$ the SQUID sensitivity (in $\mathrm{\mu A/\phi_0}$). The best sensitivity $S_\mathrm{CCC}^\mathrm{opt}$ is obtained for $N_\mathrm{PC}^\mathrm{opt}$ given by:\\
\begin{equation}
N_\mathrm{PC}^\mathrm{opt}=\sqrt{L_i/L^\mathrm{eff}_\mathrm{CCC}},
\end{equation}\\
where $L^\mathrm{eff}_\mathrm{CCC}$ is the effective self inductance of the CCC taking into account the proximity of the superconducting screen that isolates the SQUID from external magnetic fields (Earth$\mathrm{'}$s field for instance). $L^\mathrm{eff}_\mathrm{CCC}$ and therefore $S_\mathrm{CCC}$ can be determined in a given geometry using the analytical calculation of Ses\'e and co-authors\cite{Sese1999,Sese2003}. In our case, one calculates $L^\mathrm{eff}_\mathrm{CCC}\sim 14$ nH, $N_\mathrm{PC}^\mathrm{opt}=12$ and $S_\mathrm{CCC}^\mathrm{opt}=\mathrm{5~\mu A.t/\phi_0}$. Due to geometrical constraints, the number of turns of the pick-up coil was reduced to $N_\mathrm{PC}=6$ leading to an experimental sensitivity $S_\mathrm{CCC}=\mathrm{8~\mu A.t/\phi_0}$ close to the calculated value of $\mathrm{6~\mu A.t/\phi_0}$.
\section{Current sources}
\subsection{Implementation of the electronic circuits}
\label{annexe:CurrentSourcesImplementation}
\begin{figure}[h!]
\centering
\includegraphics[width=3.0in]{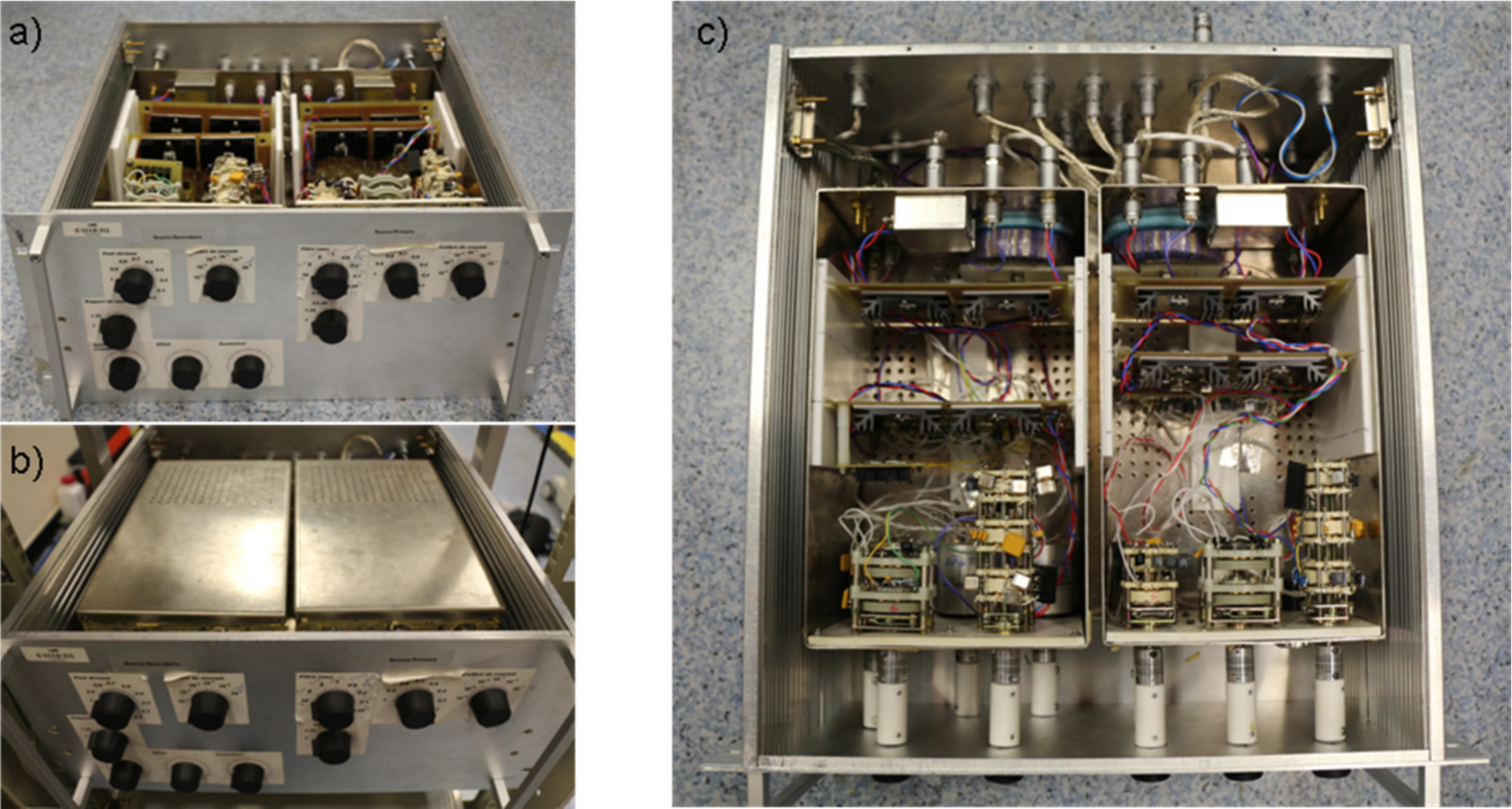}
\caption{Pictures (front a) and b), top c)) of the primary and secondary current sources, each one being placed in an independent box.}\label{FigAppendix-Sources}
\end{figure}
Electronic circuits of each current source are integrated, but electrically isolated with PTFE material, into their own metallic box connected to ground, as shown in pictures of fig.\ref{FigAppendix-Sources}. The electronic components are powered by stabilized voltages provided by a circuit itself energized by rechargeable batteries placed in their own metallic box (see fig.\ref{Fig-FullBridge}).

The only electrical link existing between the electronic circuits and the ground comes from the insulation differential amplifiers. Those which are used to probe the external piloting voltage and the SQUID feedback voltage are based on OPA128LM precision operational amplifiers (in combination with OPA97 devices) which ensure a high-isolation resistance (in principle $\sim10^{15} ~\Omega$). All these precautions aim at cancelling leakage currents. Let us note that these amplifiers are characterized by a large voltage noise at low frequencies ($\mathrm{4~\mu V_{p-p}}$ between 0.1 and 10 Hz) but the current ratio is not sensitive to it by construction. On the other hand, it is sensitive to the noise of the insulation differential amplifier at the entry of the secondary current source used to probe the voltage after the low-pass filter of the primary current source. Here, precision OPA111BM operational amplifiers with a lower voltage noise ($\mathrm{1.2~\mu V_{p-p}}$ between 0.1 and 10 Hz) are used to optimize the current ratio stability at the expense of a reduction of the isolation resistance (in principle $\sim10^{13}~\Omega$).
\subsection{Passive components of the secondary current circuit}
\label{annexe:PassiveComponents}
\begin{table}[h]
\newcolumntype{M}[1]{>{\centering\arraybackslash}m{#1}}
\begin{center}
\begin{tabular}{|c|c|c|c|c|}
  \hline
  \textbf{Range}&$\bm{R_\mathrm{C}}$&$\bm{R_\mathrm{E}}$&$\bm{C_\mathrm{F}}$&$\bm{R_\mathrm{F}}$ \tabularnewline \hline
  1 $\mathrm{\mu A}$&5 M$\Omega$&1 M$\Omega$&300 pF&4 M$\Omega$ \tabularnewline \hline
  10 $\mathrm{\mu A}$&500 k$\Omega$&100 k$\Omega$&300 pF&400 k$\Omega$  \tabularnewline \hline
  100 $\mathrm{\mu A}$&50 k$\Omega$&10 k$\Omega$&3 nF&40 k$\Omega$\tabularnewline \hline
  1 mA&5 k$\Omega$&1 k$\Omega$&30 nF&4 k$\Omega$\tabularnewline \hline
  10 mA&500 $\Omega$&100 $\Omega$&300 nF&400 $\Omega$\tabularnewline \hline
  100 mA&50 $\Omega$&10 $\Omega$&3 $\mathrm{\mu F}$&40 $\Omega$\\ \hline
\end{tabular}
\end{center}
\caption{Resistance and capacitance values in the last stage of the secondary current source for the different current ranges.}
\label{tableau:Range}
\end{table}
\section{Current dividers}
\subsection{SCD Calibration}
\label{annexe:SCD}
\begin{figure}[h!]
\centering
\includegraphics[width=3.0in]{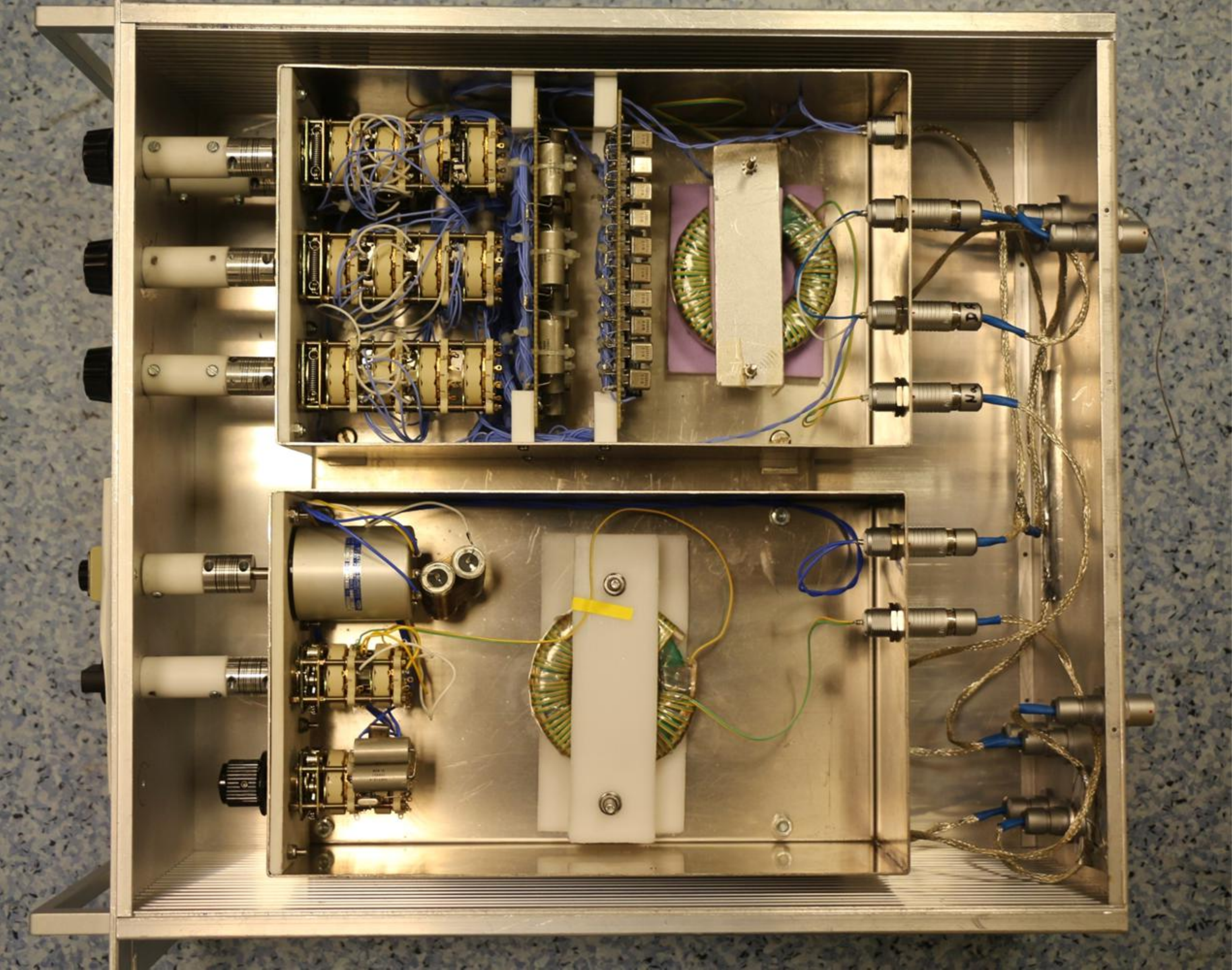}
\caption{Pictures of the standard current divider (top) and of the quadrature current divider (bottom), each one being placed in an independent box.}\label{FigAppendix-SCD}
\end{figure}
A fraction of the SCD is calibrated from the measurement of the ratio $V_\mathrm{ab}/V_{ref}$, where $V_\mathrm{ab}$ is the low voltage, indicated in fig.\ref{fig-SCD} and measured using a HP34420A nanovoltmeter (1 mV range), $V_{ref}$ is the reference voltage of a 10 V Zener standard which is applied in place of the auxiliary winding $N_\mathrm{A}$. To reduce the measurement uncertainty, the range of the HP34420A nanovoltmeter used (1 mV) is calibrated from the $V_{ref}$ voltage reference just before and after the $V_\mathrm{ab}/V_{ref}$ measurements. This is achieved by generating a calibrated 1 mV voltage from $V_{ref}=10$ V using series and parallel implementation of two ESI SR1010 transfer resistance standards (made of 1 k$\Omega$ and 10 $\Omega$ resistors respectively). The excellent linearity of the HP34420A nanovoltmeter allows the calibration of 1 mV range by the measurement of the 1 mV value only. This method allows the calibration of any fraction between 0 and $5\times10^{-5}$ with an upper bound uncertainty of $0.3\times10^{-9}$.
\subsection{QCD calculations}
\label{appendix:QCD}
Let us consider stray capacitances $C_\mathrm{P}$ and $C_\mathrm{S}$ in parallel to the resistors $R_\mathrm{P}$ and $R_\mathrm{S}$. The voltage balance equation is given by:
\begin{equation}
\frac{R_\mathrm{P}}{1+jR_\mathrm{P}C_\mathrm{P}\omega}I_\mathrm{P}-\frac{R_\mathrm{S}}{1+jR_\mathrm{S}C_\mathrm{S}\omega}I_\mathrm{S}=0
\end{equation}
Using the ampere.turn balance equation \ref{Equation:AmpereTurnsQCD} leads to:
\begin{multline}
[R_\mathrm{P}(N_\mathrm{S}+\epsilon_\mathrm{eq}N_\mathrm{A})-R_\mathrm{S}N_\mathrm{P}-R_\mathrm{P}R_\mathrm{S}\epsilon_\mathrm{eq}^q N_\mathrm{A}^q\omega^2]\\
+j\omega R_\mathrm{P}[R_\mathrm{S}(N_\mathrm{S}+\epsilon_\mathrm{eq}N_\mathrm{A})-R_\mathrm{S}N_\mathrm{P}C_\mathrm{P}-\epsilon_\mathrm{eq}^qN_\mathrm{A}^q]=0
\end{multline}
First-order approximation in $\omega$ applied to both the real and imaginary parts of the equation leads to equation \ref{Equation:resistance} and equation \ref{Equation:Phase}.
\section{Estimation of the SQUID feedback accuracy}
\label{annexe:CLG}
Although the SQUID amplifier is based on an integrator which leads to an infinite gain at dc, measurements are in practice carried out with a finite time periodicity of the current reversal, typically of 70 s. This causes a relative current ratio error, which is equal to $(\Delta^{Prior-adj}r_I/r_I)\times(G_\mathrm{CLG}/G_\mathrm{OLG})$, where $\Delta^{Prior-adj}r_I/r_I$ is the relative deviation between the target ratio and the preliminary current ratio as it is adjusted prior to the SQUID feedback operation. Measurements of the $100~\Omega/(R_\mathrm{K}/2)$ resistance ratio were performed for $\Delta^{Prior-adj}r_I/r_I$ values as large as a few $\pm10^{-4}$. From the discrepancies measured, a relative error in the measurement of the resistance ratio lower than $10^{-11}$ is deduced for the usual adjustment of the current ratio, i.e. for $\Delta^{Prior-adj}r_I/r_I\sim 2\times10^{-6}$ (as discussed in subsection \ref{subsection: Adjustability}). This corresponds to a value $G_\mathrm{OLG}>7.5\times10^{4}$ V/$\phi_0$, compatible with that determined in quantized current experiment\cite{Brun-Picard2016}.
\section*{Acknowledgment}
The authors would like to thank Guillaume Spengler and Laetitia Soukiassian for their work during the first stage of the development of the resistance measurement bridge that started ten years ago as well as Fran\c{c}ois Piquemal for useful comments on the manuscript. The authors would like also to thank Carlos Sanchez from NRC for useful discussions about the resistance bridge operation with a ground connected to the secondary winding.
\ifCLASSOPTIONcaptionsoff
  %\newpage
\fi



\bibliographystyle{IEEEtran}
\providecommand{\noopsort}[1]{}\providecommand{\singleletter}[1]{#1}%

%\begin{thebibliography}{1}


%\end{thebibliography}

\begin{IEEEbiography}[{\includegraphics[width=1in,height=1.25in,clip,keepaspectratio]{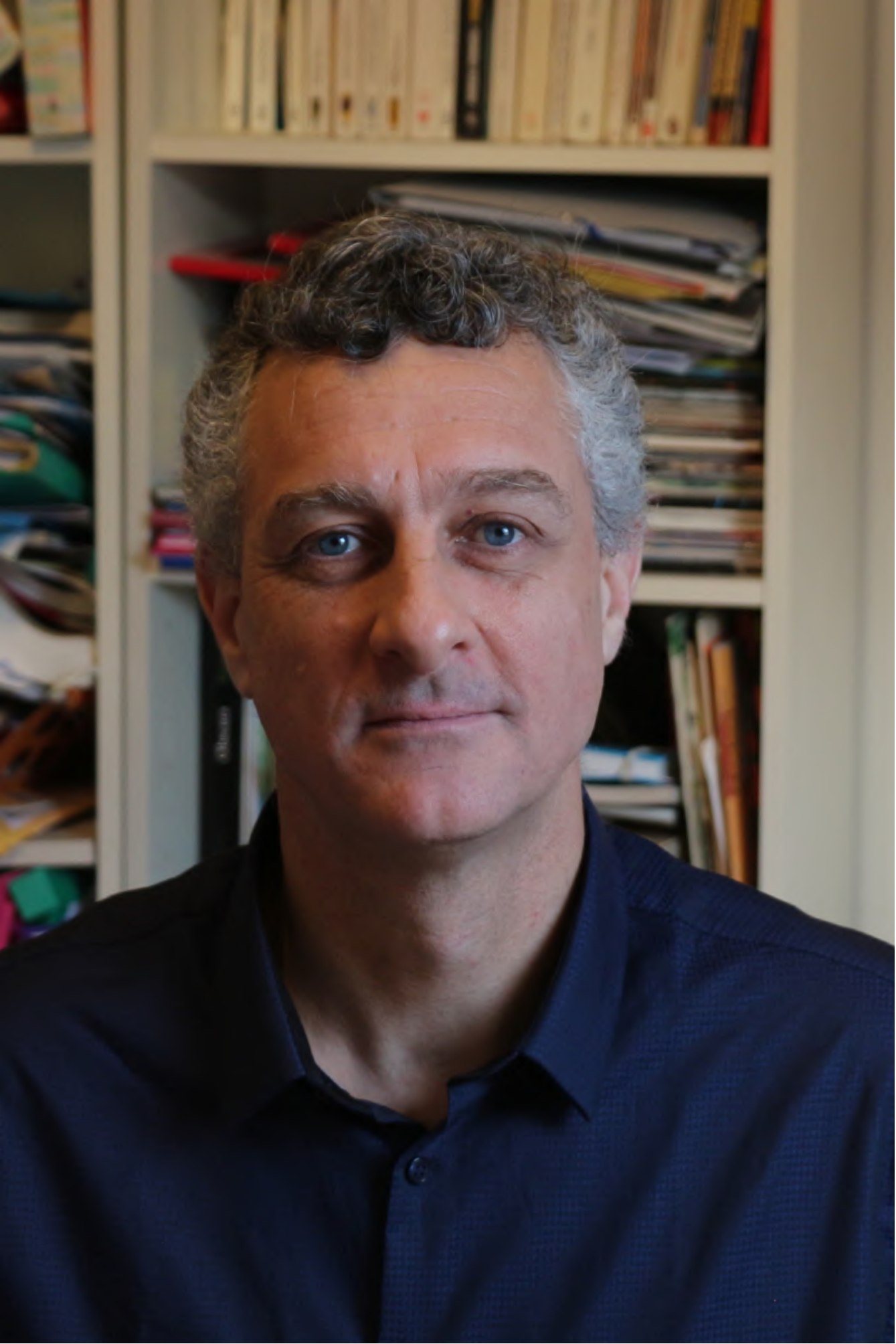}}]{Wilfrid Poirier}
is a  graduate of the Ecole Sup\'erieure de Physique et de Chimie Industrielles de Paris (ESPCI). He then completed his thesis on quantum electronic transport at the CEA-SPEC and received his doctorate in solid state physics in 1997. He joined LCIE in 1998 and then LNE in 2001 as head of studies on quantum resistance standards. Since then, he has devoted his research to quantum electrical metrology. These included the development of a graphene quantum resistance standard, of GaAs-based quantum Hall arrays and the realization of universality tests of the quantum Hall effect. He has also been involved in the development of precision quantum instrumentation based on SQUID technology. More recently, he has proposed and developed a quantum current generator to achieve the new definition of ampere. He obtained his Habilitation to Direct Research from the University of Paris-SUD in 2017.
\end{IEEEbiography}
\begin{IEEEbiography}[{\includegraphics[width=1in,height=1.25in,clip,keepaspectratio]{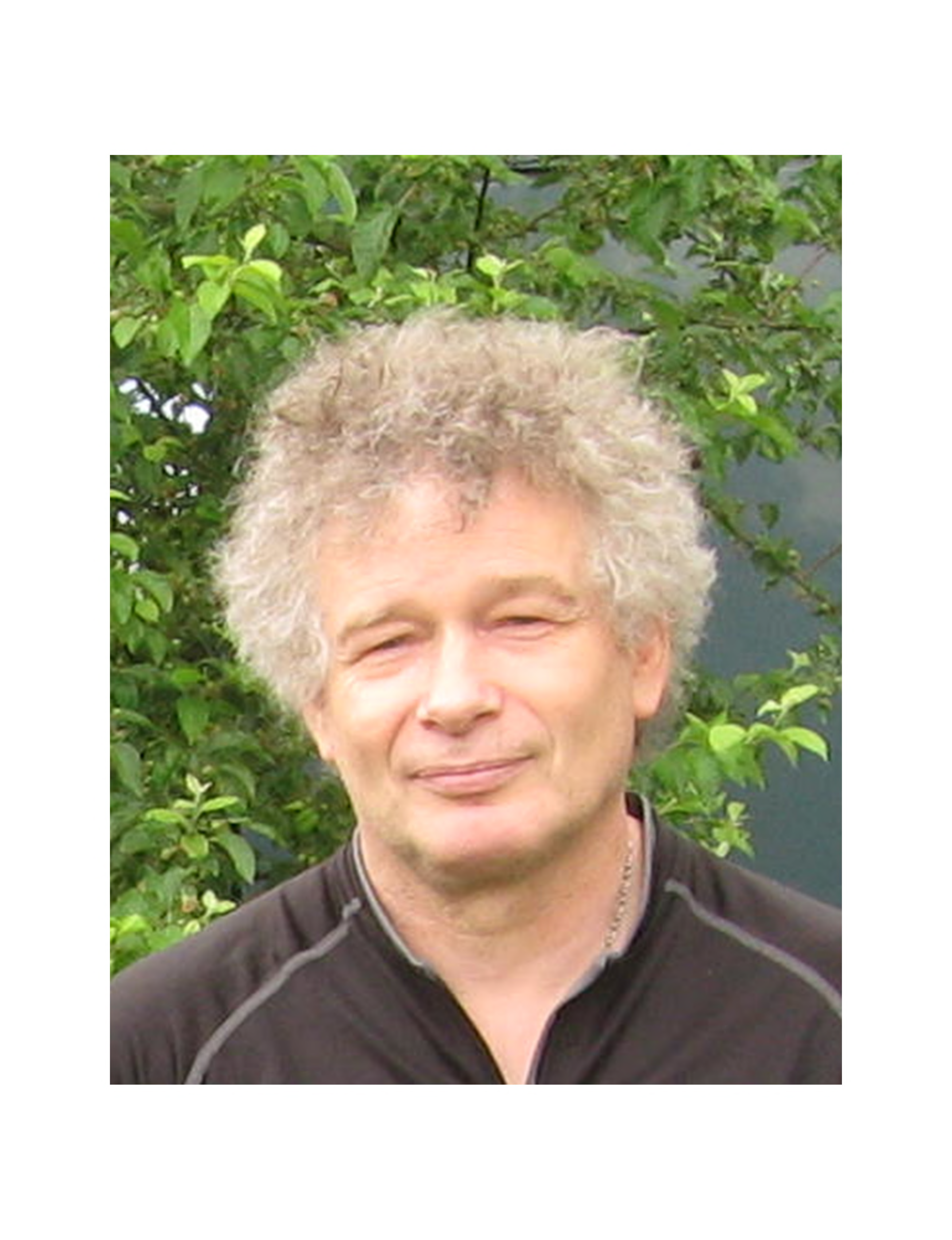}}]{Dominique Leprat}
worked in LNE from 2001 to 2017 as a technician in the electrical metrology department. He was first involved in the activity of traceability of alternating voltage based on thermal transfer. In 2007, he joined the quantum Hall effect team to develop metrology instrumentation.
\end{IEEEbiography}
\begin{IEEEbiography}[{\includegraphics[width=1in,height=1.25in,clip,keepaspectratio]{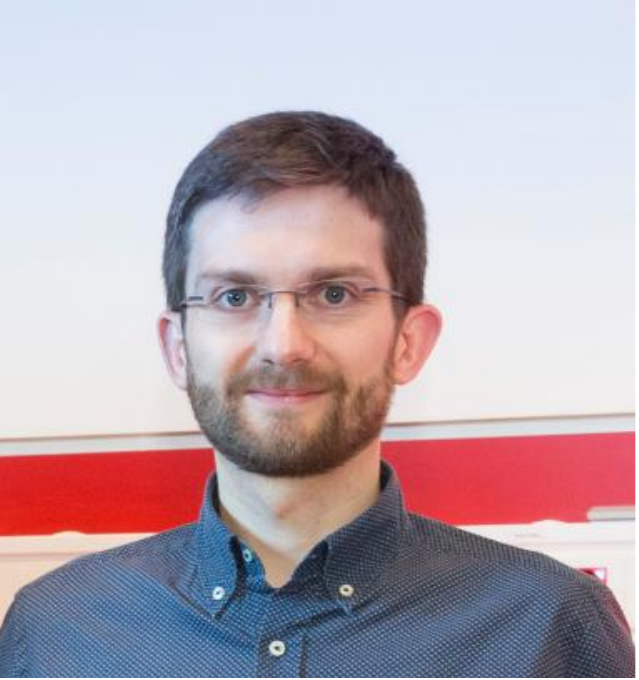}}]{F\'elicien Schopfer}
obtained an engineer's degree from \'Ecole Nationale Sup\'erieure de Physique - Grenoble INP in 2001, a master's degree in condensed matter physics from the University of Grenoble the same year. He received a PhD in Physics from the University of Grenoble in 2005, for quantum electronic transport experiments in nanostructures carried out at CNRS. He was appointed at Laboratoire national de métrologie et d’essais – LNE in 2005 to advance research in quantum electrical metrology. His research has mainly focused on the quantum Hall effect (QHE) for applications in fundamental metrology. He worked on quantum Hall arrays in GaAs/AlGaAs, co-authored reproducibility and universality tests of the quantum Hall effect with record uncertainties, and is strongly involved in graphene research, notably with important results for the development of the quantum Hall resistance standard operating under relaxed experimental conditions.
\end{IEEEbiography}
\end{document}